\pdfoutput=1
\documentclass[
    10pt,
    final,
    journal,
    comsoc,
    letterpaper,
    twocolumn,
]{IEEEtran}

\ifCLASSOPTIONtwocolumn%
    \usepackage{balance}
\fi

\usepackage[T1]{fontenc}
\usepackage{rotating}                   
\usepackage{graphicx}                   

\ifCLASSOPTIONcompsoc%
    \usepackage[
        caption=false,
        font=normalsize,
        labelfont=sf,
        textfont=sf
    ]{subfig}
\else
    \usepackage[
        caption=false,
        font=footnotesize
    ]{subfig}
\fi

\usepackage{siunitx}                    
\usepackage[cmex10,intlimits]{amsmath}  
\usepackage{cite}                       
\usepackage{amsthm}                     
\usepackage{amssymb}                    
\usepackage{bm}                         
\usepackage[
    bookmarks=true,
    bookmarksnumbered=true,
    pdfpagemode=UseOutlines,
    plainpages=false,
    pdfpagelabels=true,
    colorlinks=true,
    allcolors=black,
    pagebackref=false,
    pdfkeywords={{Indoor environments, Algorithm design and analysis, Estimation,Optimization, Bayes methods}},
    pdftitle={A Unifying Bayesian Optimization Framework for Radio Frequency Localization},
    pdfauthor={Nachikethas A. Jagadeesan},
]{hyperref}                             
\usepackage{thumbpdf}
\usepackage[all]{hypcap}                
\usepackage[
    final,
    tracking=true,
    kerning=true,
    spacing=true,
    factor=1100
]{microtype}                            
\usepackage{booktabs}                   
\usepackage[inline]{enumitem}
\usepackage{array}                      

\newcolumntype{L}[1]{>{\raggedright\let\newline\\\arraybackslash\hspace{0pt}}m{#1}}
\newcolumntype{C}[1]{>{\centering\let\newline\\\arraybackslash\hspace{0pt}}m{#1}}
\newcolumntype{R}[1]{>{\raggedleft\let\newline\\\arraybackslash\hspace{0pt}}m{#1}}

\newcommand{\cv}{{\bm c}}

\newcommand{\lv}{{\bm l}}

\newcommand{\ov}{{\bm o}}

\newcommand{\rv}{{\bm r}}

\newcommand{\xv}{{\bm x}}

\newcommand{\zerov}{{\bm 0}}

\newcommand{\Ov}{{\bm O}}

\newcommand{\Rv}{{\bm R}}

\DeclareMathOperator*{\argmax}{arg\,max}
\DeclareMathOperator*{\argmin}{arg\,min}

\DeclareMathOperator*{\E}{\mathrm{E}}

\newcommand{\interior}[1]{%
  {\kern0pt#1}^{\mathrm{o}}%
}

\newtheorem{thm}{Theorem}

\newtheorem*{cor*}{Corollary}

\newtheorem{con}{Conjecture}

\theoremstyle{remark}

\newtheorem*{rem*}{Remark}

\theoremstyle{definition}
\newtheorem{dfn}{Definition}

\DeclareSIUnit\dbm{\decibel{}m}
\DeclareSIUnit\db{\decibel}

\newcolumntype{L}[1]{>{\raggedright\let\newline\\\arraybackslash\hspace{0pt}}m{#1}}
\newcolumntype{C}[1]{>{\centering\let\newline\\\arraybackslash\hspace{0pt}}m{#1}}
\newcolumntype{R}[1]{>{\raggedleft\let\newline\\\arraybackslash\hspace{0pt}}m{#1}}

\begin{document}

\title{A Unifying Bayesian Optimization Framework for Radio Frequency Localization}

\author{Nachikethas~A.~Jagadeesan and~Bhaskar~Krishnamachari%
\thanks{The authors are with the Department of Electrical Engineering,
University of Southern California, Los Angeles, CA 90089.
Email: \texttt{\{nanantha, bkrishna\}@usc.edu}
}
}

\maketitle

\begin{abstract}
    We consider the problem of estimating an RF-device's location based on
observations, such as received signal strength, from a set of
transmitters with known locations. We survey the literature on this
problem, showing that previous authors have considered implicitly or
explicitly various metrics.  We present a novel Bayesian framework that
unifies these works and shows how to optimize the location estimation
with respect to a given metric. We demonstrate how the framework can
incorporate a general class of algorithms, including both model-based
methods and data-driven algorithms such fingerprinting. This is
illustrated by re-deriving the most popular algorithms within this
framework. When used with a data-driven approach, our framework has
cognitive self-improving properties in that it provably improves with
increasing data compared to traditional methods. Furthermore, we propose
using the error-CDF as a unified way of comparing algorithms based on
two methods: (i) stochastic dominance, and (ii) an upper bound on
error-CDFs. We prove that an algorithm that optimizes any distance based
cost function is not stochastically dominated by any other algorithm.
This suggests that in lieu of the search for a universally best
localization algorithm, the community should focus on finding the best
algorithm for a given well-defined objective.

\end{abstract}


\section{Introduction}~\label{sec:intro}
\IEEEPARstart{T}{he} ability to locate a wireless device based on received signal
strength from known-location transmitters is of great utility for many
applications, including indoor location-based mobile apps and services,
interactive media, emergency search and rescue, asset tracking, etc. A
significant number of researchers have tackled this fundamental problem
and proposed various algorithms for radio signal strength (RSS) based
localization. Many works adopt standard algorithms from signal
processing, specifically estimation theory, such as Maximum Likelihood
Estimation~\cite{waa10}, Minimum Mean Squared Error
Estimation~\cite{hua10}, Best Linear Unbiased Estimator~\cite{lin11},
etc., while other techniques such as finger-printing~\cite{bah00}, and
sequence-based localization~\cite{yed08}, are somewhat more
heuristically derived. These algorithms are typically evaluated using
numerical and trace-based simulations, using varied metrics such as the
mean squared position error, the absolute distance error, etc.

We contend that the literature is disconnected and disorganized and that
it is hard to decipher any unified theory that fairly evaluates these
algorithms across different metrics of interest. We argue that the
state-of-the-art approach to localization in the literature --- which
typically involves first presenting an algorithm and then evaluating its
performance according to a particular metric or a set of metrics --- is
akin to putting the proverbial cart before the horse. For instance, it
is not uncommon for algorithms to be evaluated on metrics for which they
are not explicitly or implicitly optimized.

We advocate a systematic way of designing location estimation algorithms
which we refer to as the ``optimization-based approach to
localization''.
In this approach, first the localization metric is defined in terms of a
suitable cost function, then an appropriate estimation algorithm is
derived to minimize that cost function. In addition, our optimization
framework is applicable to any deterministic or stochastic model of the
observations (assumed to be known) and can accommodate any prior
distribution for location. Our framework also applies to data-driven
approaches such as fingerprinting. We show that, in such data-driven
settings, our framework
\begin{enumerate*}[label = (\roman*)]
    \item makes better use of the available data, and
    \item has cognitive self-improving properties in that it provably
        gets better with increasing data
\end{enumerate*}
compared to traditional methods.
Fundamentally a Bayesian approach, this
framework is also compatible with Bayesian filtering for location
tracking over time~\cite{fox03}.

As an illustration of our framework, we consider first a common metric
used in the evaluation of localization algorithms, the absolute distance
error, and derive an algorithm which yields location estimates so as to
minimize the expected distance error (MEDE). For a second illustration,
we also consider as another metric the probability that the location
estimate is within a given radius of the true location ($P(d)$) and
derive an algorithm which maximizes this probability.
Furthermore, we
show that standard algorithms such as MLE and MMSE can be derived
similarly from optimizing the corresponding metrics (likelihood,  mean
squared error respectively).

In conjunction with our framework for deriving algorithms to optimize a
specified metric, we also consider the problem of comparing different
localization algorithms with each other; for which we propose the use of
the error CDF\@. For an important class of cost
functions that can be expressed as non-negative monotonically increasing
functions of distance error, we prove that there is in effect a partial
ordering among various estimation algorithms. Certain algorithms
dominate other algorithms for all such cost functions, and we show the
necessary condition for this to happen. But there could also be two
algorithms, $A_1$ and $A_2$, and two metrics, $M_1$ and $M_2$, such that
$A_1$ is better than $A_2$ in terms of $M_1$ while the reverse is true
for $M_2$. Thus we show that there is, in general, no single-best
localization algorithm, but rather a ``Pareto Set'' of algorithms that
are each optimized for different cost functions.

We evaluate the optimization-based framework for location estimation
using both numerical simulations, traces~\cite{bau09}, and data obtained
from experiments in an indoor office environment. We illustrate how our
framework can incorporate a variety of localization algorithms,
including fingerprinting based methods.
Our evaluation
confirms what is predicted by the theory --- no single algorithm
outperforms others with respect to all metrics, thus underlining the
need for an optimization based apporach such as the one we propose.

\subsection{Contributions}
\begin{itemize}
    \item We propose a novel optimization-based Bayesian framework that
        unifies and puts in context various previously proposed
        techniques for localization and provides a systematic basis for
        developing new algorithms.
    \item We demonstrate the usage of error CDF as a unifying
        mathematical abstraction for evaluating the performance of a
        localization algorithm. We first introduce a partial ordering
        over the set of algorithms by considering a stochastic dominance
        relationship between their error CDFs. We prove that any
        algorithm that optimizes a distance based cost function is not
        stochastically dominated by any other algorithm.
    \item We further demonstrate a second way of evaluating and
        comparing algorithms based on how `close' an algorithm gets to
        the upper bound on error CDFs. We propose one such measure of
        closeness (area of difference) and identify MEDE as the optimal
        algorithm over that measure.
    \item We illustrate how our framework encompasses both model-based
        approaches and data-driven methods such as fingerprinting,
        through simulations and real-world experiments.
\end{itemize}

The rest of the paper is organized as follows: A survey of the existing
literature on RSS based localization algorithms is given in
Section~\ref{sec:related}. We introduce our optimization based
localization framework in Section~\ref{sec:framework}. In
Section~\ref{sec:errcdf} we introduce the concept of stochastic
dominance and prove how it leads to a partial ordering over the set of
localization algorithms. We also introduce the upper bound on error CDFs
and illustrate the evaluation of algorithms using the same. In
Section~\ref{sec:evaluation} we evaluate our framework using simulations
and trace data. We also show how fingerprinting methods fit into
our framework. We conclude in
Section~\ref{sec:conclusion}.

\section{Literature on RSS based Localization}~\label{sec:related}
\begin{table}[!ht] \centering
\renewcommand{\arraystretch}{1.25}
\caption{Literature Survey on Localization Methods}
\label{tab:litsur}
 \begin{tabular}{@{} L{0.14\textwidth} L{0.07\textwidth} l l @{}} \toprule
    \emph{Study} & \emph{Algorithm} & \emph{Model} & \emph{Metric}\\
    \midrule
    WLAN location determination via clustering and probability
    distributions~\cite{you03} & MLE & Fingerprinting & $P(d)$\\
    \midrule
    Maximum likelihood localization estimation based on received signal
    strength~\cite{waa10} & MLE & Log-Normal & MSE \\
    \midrule
    Experimental comparison of RSSI-based localization algorithms for indoor wireless
    sensor networks~\cite{zan08} & Multilateration and MLE & Log-Normal & EDE\\
    \midrule
    A Bayesian sampling approach to in-door localization of wireless devices using
    RSSI~\cite{ses05} & Weighted MLE and Error CDF & Fingerprinting & EDE \\
    \midrule
    RADAR\@: An In-Building RF-based user location and tracking system~\cite{bah00} &
    Clustering & Fingerprinting & EDE \\
    \midrule
    The Horus WLAN location determination system~\cite{you05} & MLE & Fingerprinting
    & EDE \\
    \midrule
    Locating in fingerprint space: Wireless indoor localization with little human 
    intervention~\cite{yan12} & Clustering & Fingerprinting & EDE \\
    \midrule
    Weighted centriod localization in Zigbee-based sensor networks~\cite{blu07} &
    RSSI weighted position & Free Space Path Loss & EDE \\
    \midrule
    Sequence based localization in wireless sensor networks~\cite{yed08} & SBL &
    Free Space Path Loss & EDE \\
    \midrule
    Best linear unbiased estimator algorithm for RSS based localization~\cite{lin11} &
    Best linear unbiased estimate & Log-Normal & MSE \\
    \midrule
    Performance of an MMSE based indoor localization with wireless sensor
    networks~\cite{hua10} & MMSE & Log-Normal & MSE \\
    \midrule
    Relative location estimation in wireless sensor networks~\cite{pat03} & MLE &
    Log-Normal & MSE\\
    \midrule
    Distance Estimation from RSS under log-normal shadowing~\cite{chi09.1} & Best
    unbiased estimate & Log-Normal & MSE \\
    \bottomrule
\end{tabular}
\end{table}

In this section, we survey the existing literature on RSS based localization
algorithms with the intention of comparing, across papers, the metrics used to
evaluate the algorithms. The results of the survey are summarized in
Table~\ref{tab:litsur}. We identify certain metrics that are commonly used
across the literature:
\begin{itemize}
    \item \emph{MSE}: Mean Squared Error is the expected value of the
        square of the Euclidean distance between our estimate and the
        true location. Often the square root of this quantity Root Mean
        Squared Error (RMSE) is given instead of MSE\@. As RMSE may be
        derived using MSE, we shall only use MSE in our discussions in
        this paper. The minimum mean squared error (MMSE) algorithm
        returns an estimate that minimizes the MSE\@.
    \item \emph{EDE}: The Expected Distance Error (EDE) is the expected
        value of the Euclidean distance between our estimate and the
        true location. The minimum expected distance error (MEDE)
        algorithm returns an estimate that minimizes the EDE\@.
    \item \textbf{$P(d)$}: $P(d)$ indicates the probability that the
        receiver location is within a distance of $d$ from our location
        estimate. $P(d)$ is closely related to the metric $D(p)$ which
        gives the radius at which an open ball around our location
        estimate yields a probability of at least $p$. The M$P(d)$
        algorithm returns an estimate that minimizes the $P(d)$.
\end{itemize}
As evidenced by Table~\ref{tab:litsur}, it is with striking regularity
that one encounters a mismatch between an algorithm and the metric used
for its evaluation. While there is hardly anything amiss in checking how
an algorithm performs on a metric that it is not optimized for, it is
shortsighted to draw a conclusion as to the efficacy of the said
algorithm based on such an evaluation. An awareness of the metric that
an algorithm is implicitly or explicitly optimized for, is essential to
its fair assessment. We believe that such of notion of \emph{consistent}
evaluation of algorithms across \emph{all} important metrics of interest
has been absent in the community so far. In addition while the
literature on localization abounds in algorithms that yield a location
estimate, there is no unifying theory that relates them to each other
with appropriate context.

For instance,~\cite{aks11}~picks four algorithms for evaluation,
indepedent of the metrics used to evaluate the algorithms. Such an
approach makes it unclear if an algorithm is optimal with respect with
any of the given metrics. In this case, we can only make (empirical)
inferences regarding the relative ordering of the \emph{chosen}
algorithms among the \emph{chosen} metrics. Consequently, there are no
theoretical guarentees on algorithm performance and it becomes hard, if
not impossible, to accurately predict how a chosen algorithm will behave
when evaluated with a metric that was not considered.

Moreover, while error CDFs have been used earlier to compare
algorithms~\cite{eln04}, they are typically used to derive inferences
about algorithm performance with respect to the Euclidean distance and
$D(p)$ metrics. In the absence of the unifying theory presented in this
proposal, it is unclear how one may draw meaningful conclusions
regarding the relative performance of algorithms across various metrics
based on their error CDF\@. Our proposed unifying framework places the
commonly employed subjective reading of error CDFs on a firm theoretical
footing and enables a better understanding of algorithm performance than
what was previously possible. Moreover, our framework is computationally
tractable as the optimization is typically done over a reasonably sized
discrete set of possible locations.

Table~\ref{tab:litsur} also indicates that there is considerable
interest in the community for the EDE metric. However, it is interesting
to note that \emph{none} of the algorithms evaluated using that metric
are explicitly optimized for it. In the following sections we show how
such metrics fit into our framework. More importantly, it is our hope
that thinking in terms of the framework below shall lead to a clearer
understanding of the trade-offs involved in choosing an algorithm and a
better specification of the criterion necessary for its adoption.

\begin{figure}[t]
    \centering
    \includegraphics[width=0.49\textwidth]{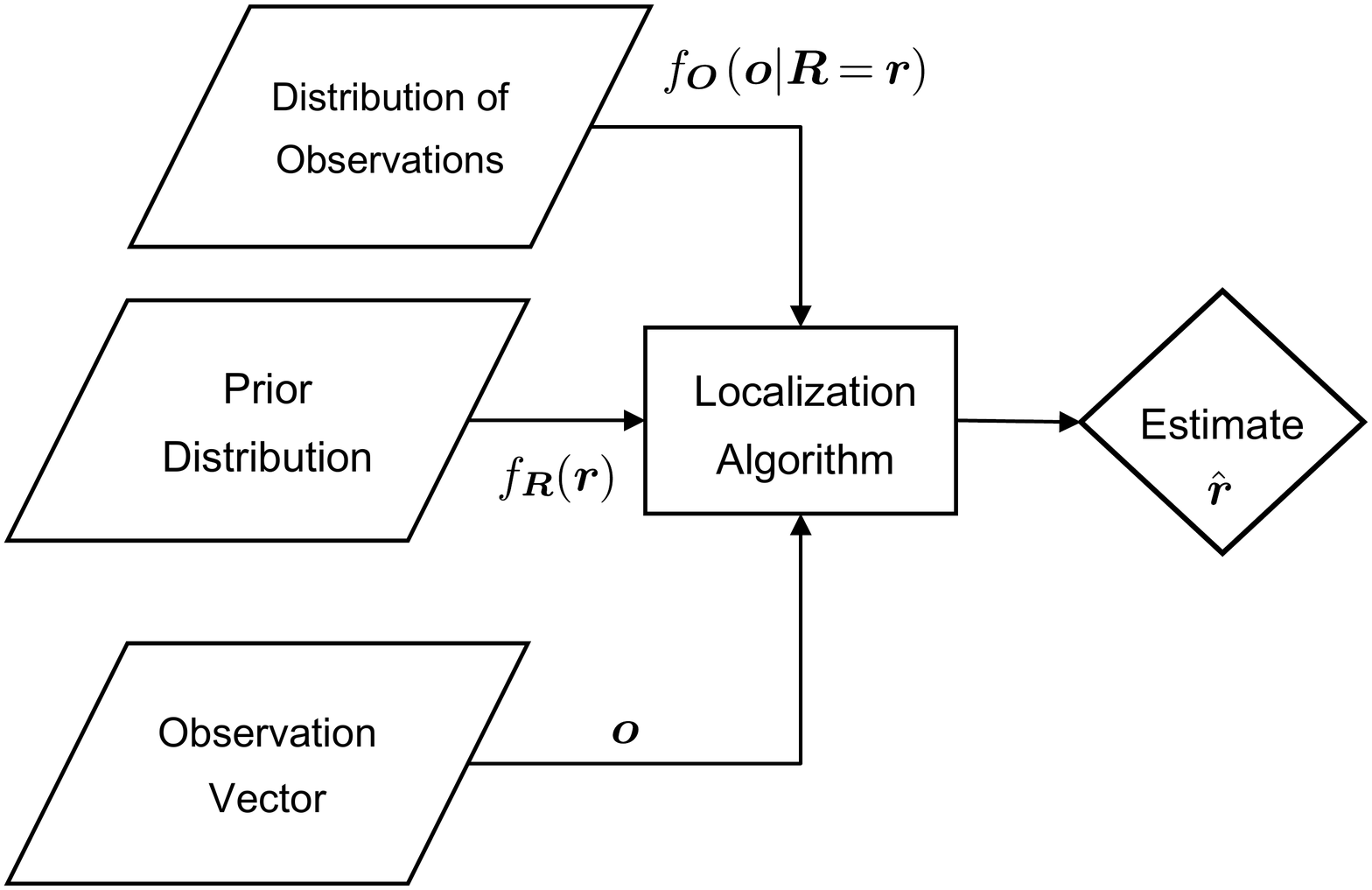}
    \caption{Localization Algorithms}\label{fig:algo}
\end{figure}

The optimization-based framework for localization presented in this
paper is inspired in part by the optimization-based approach to
networking developed since the late 90's~\cite{LowLapsley}, which has
shown successfully that efficient medium access, routing, and congestion
control algorithms, protocols, and architectures all can be derived from
suitably specified network utility maximization problems~\cite{Srikant}.
Moreover, Bayesian optimization is increasingly gaining in popularity in
the recent years~\cite{sha16}, including applications in cognitive radio
networks~\cite{jac14,xin13}, largely due to the increased availability
of both abundant data and the computational power needed to process that
data. Our proposed framework is poised to leverage both these trends.

\section{A Unifying Framework for Deriving Localization Algorithms}~\label{sec:framework}
\begin{table*}[ht] \centering
\renewcommand{\arraystretch}{1.5}
\caption{Recovering existing algorithms in our framework}
\label{tab:defalgos}%
\begin{tabular}{@{} l l l @{}} \toprule
     \emph{Algorithm} & \emph{Cost function} & \emph{Optimization}\\
     \midrule
     MMSE & $\mathcal{C}(\rv, \tilde{\rv}, \ov) =
        {\left(\| \tilde{\rv} - \rv \|_2\right)}^2$ &
        ${\rv}_{MMSE} =  \argmin_{\tilde{\rv}}
        \E\left[{\left(\| \tilde{\rv} - \rv \|_2\right)}^2\right]$\\
     MEDE & $\mathcal{C}(\rv, \tilde{\rv}, \ov) =
        \| \tilde{\rv} - \rv \|_2$ &
        ${\rv}_{MEDE} =  \argmin_{\tilde{\rv}}
        \E\left[\| \tilde{\rv} - \rv \|_2\right]$\\
     M$P(d)$ & $\mathcal{C}(\rv, \tilde{\rv}, \ov) =
        -P\left(\| \tilde{\rv} - \rv \|_2 \le d\right)$ &
        ${\rv}_{MP(d)} =  \argmax_{\tilde{\rv}}
        \E\left[P\left(\| \tilde{\rv} - \rv \|_2 \le d\right)\right]$\\
     MLE & $\mathcal{C}(\rv, \tilde{\rv}, \ov) =
        -P\left(\| \tilde{\rv} - \rv \|_2 \le \epsilon\right)$ &
        ${\rv}_{MLE} =  \lim_{\epsilon \to 0}\argmax_{\tilde{\rv}}
        \E\left[P\left(\| \tilde{\rv} - \rv \|_2 \le \epsilon\right)\right]$\\
     \bottomrule
\end{tabular}
\end{table*}

Let $\mathcal{S} \subseteq \mathbb{R}^2$ be the two-dimensional space of
interest in which localization is to be performed\footnote{It is trivial
    to extend the framework to 3-D localization, for simplicity, we
    focus on the more commonly considered case of 2-D localization
here.}. We assume that $\mathcal{S}$ is closed and bounded. Let the
location of the receiver (the node whose location is to be estimated) be
denoted as $\rv = [x_r, y_r]$. Using a Bayesian viewpoint, we assume
that this location is a random variable with some prior distribution
$f_{\Rv}(\rv)$. This prior distribution is used to represent knowledge
about the possible position, obtained, for instance from previous
location estimates or knowledge of the corresponding user's mobility
characteristics in the space; in the absence of any prior knowledge, it
could be set to be uniform over $\mathcal{S}$. Let $\ov \in
\mathbb{R}^N$ represent the location dependent observation data that was
collected. As an example, $\ov$ could represent the received signal
strength values from transmitters whose locations are known.
Mathematically, we only require that the observation vector is drawn
from a known distribution that depends on the receiver location $\rv$:
$f_{\Ov}\left( \ov \vert \Rv = \rv \right)$. In case of RSS
measurements, this distribution characterizes the stochastic radio
propagation characteristics of the environment and the location of the
transmitters. Note that this distribution could be expressed in the form
of a standard fading model whose parameters are fitted with observed
data, such as the well-known simple path loss model with log-normal
fading~\cite{mol10}.
The distribution  $f_{\Ov}\left( \ov \vert \Rv = \rv
\right)$ is general enough to incorporate more data-driven approaches such as the
well-known finger-printing procedure. In finger-printing, there is a training phase in
which statistical measurements are obtained at the receiver at various known locations and
used to estimate the distribution of received signal strengths at each
location.\footnote{We note that in many implementations of fingerprinting, only the mean
received signal strength from each transmitter is used, which of course is a special case,
equivalent to assuming a deterministic signal strength measurement with a unit step
function cumulative distribution function.} Fundamentally, the data-driven approach
constructs $f_{\Ov}\left( \ov \vert \Rv = \rv \right)$ empirically, while
model-dependent approaches take the distribution over observations directly from the model.

\begin{figure*}[!t]
    \centering
    \subfloat[]{\includegraphics[width=0.49\textwidth]
    {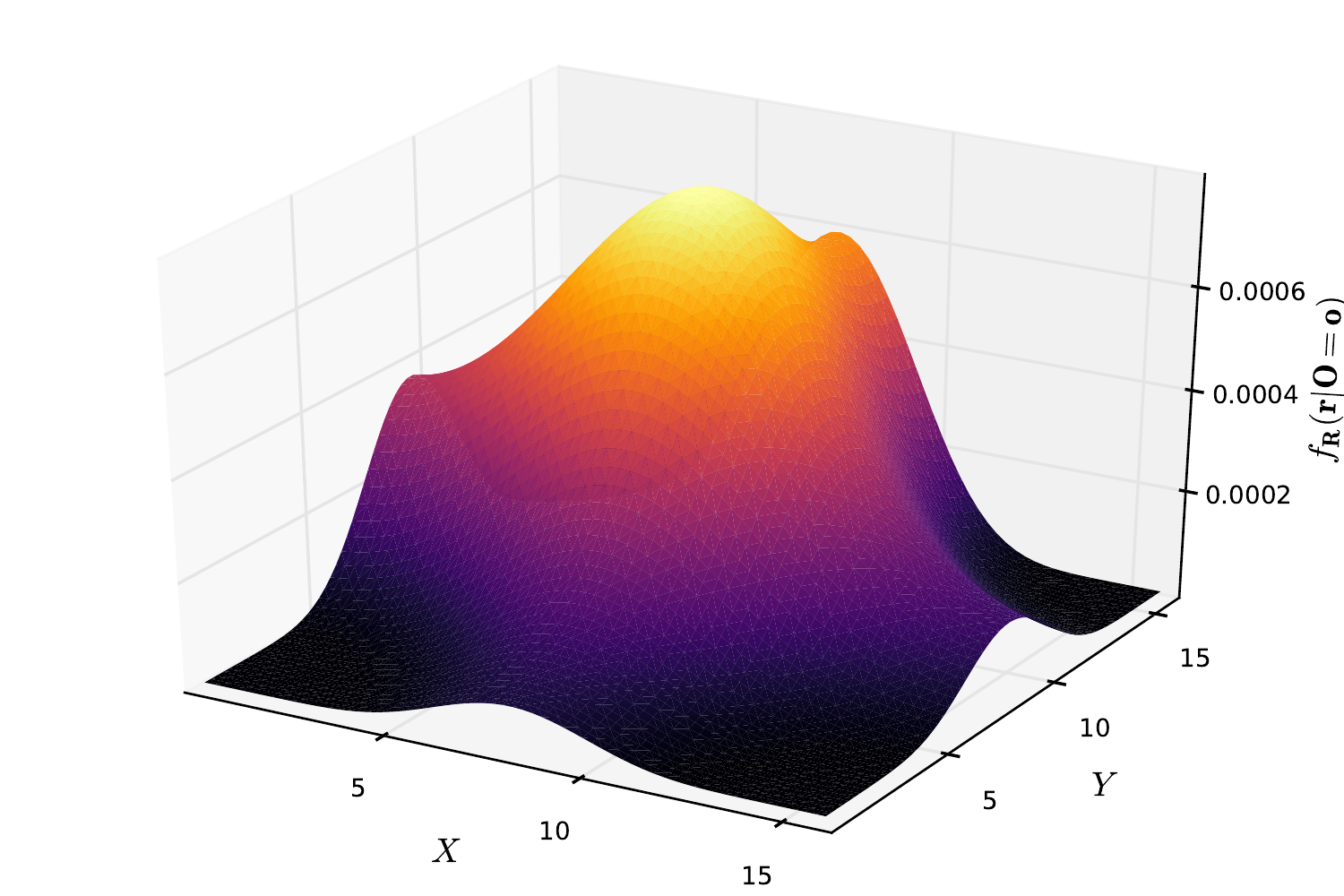}\label{fig:pdfRgO_1}}
    \hfil
    \subfloat[]{\includegraphics[width=0.49\textwidth]
    {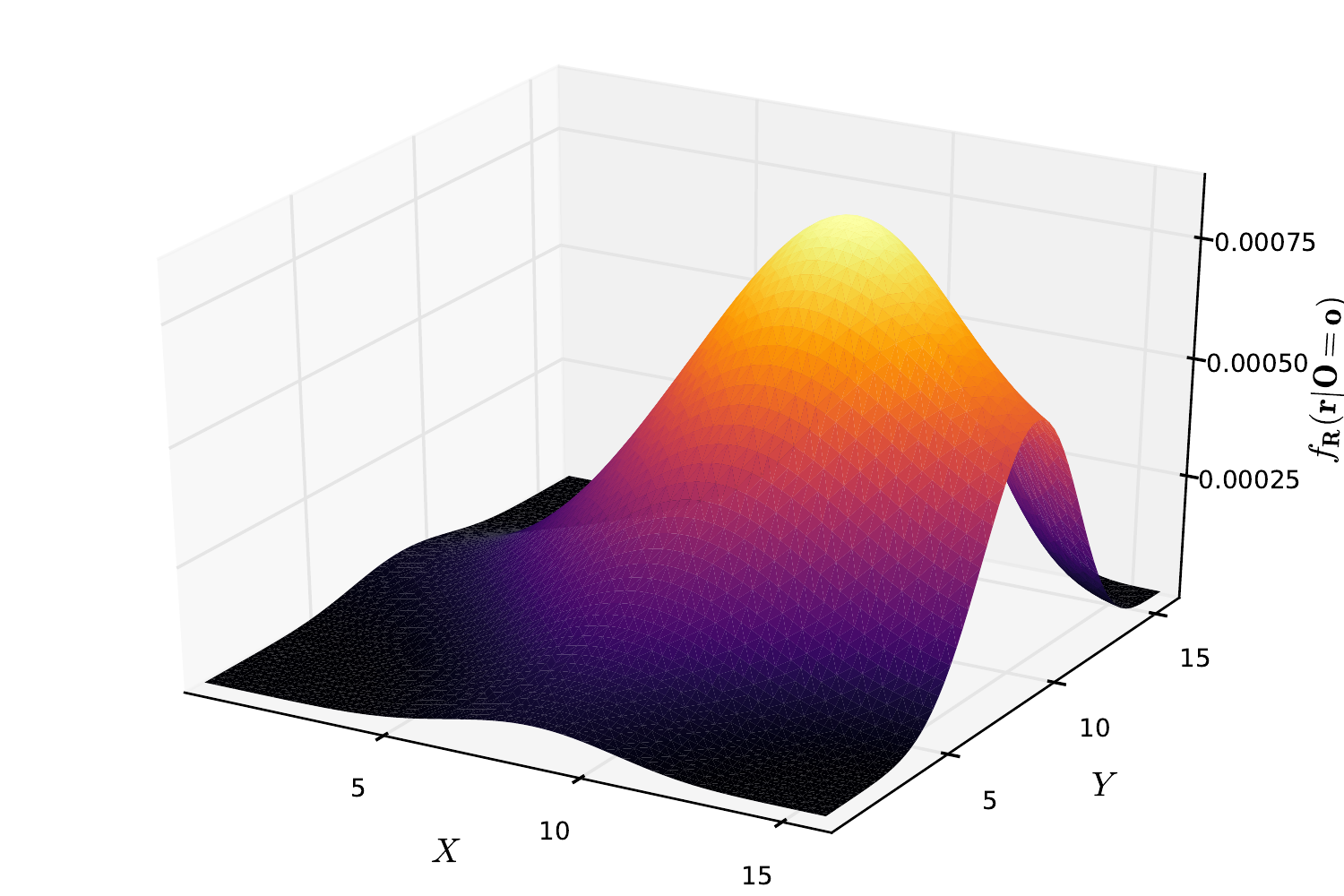}\label{fig:pdfRgO_2}}\\
    \subfloat[]{\includegraphics[width=0.49\textwidth]
    {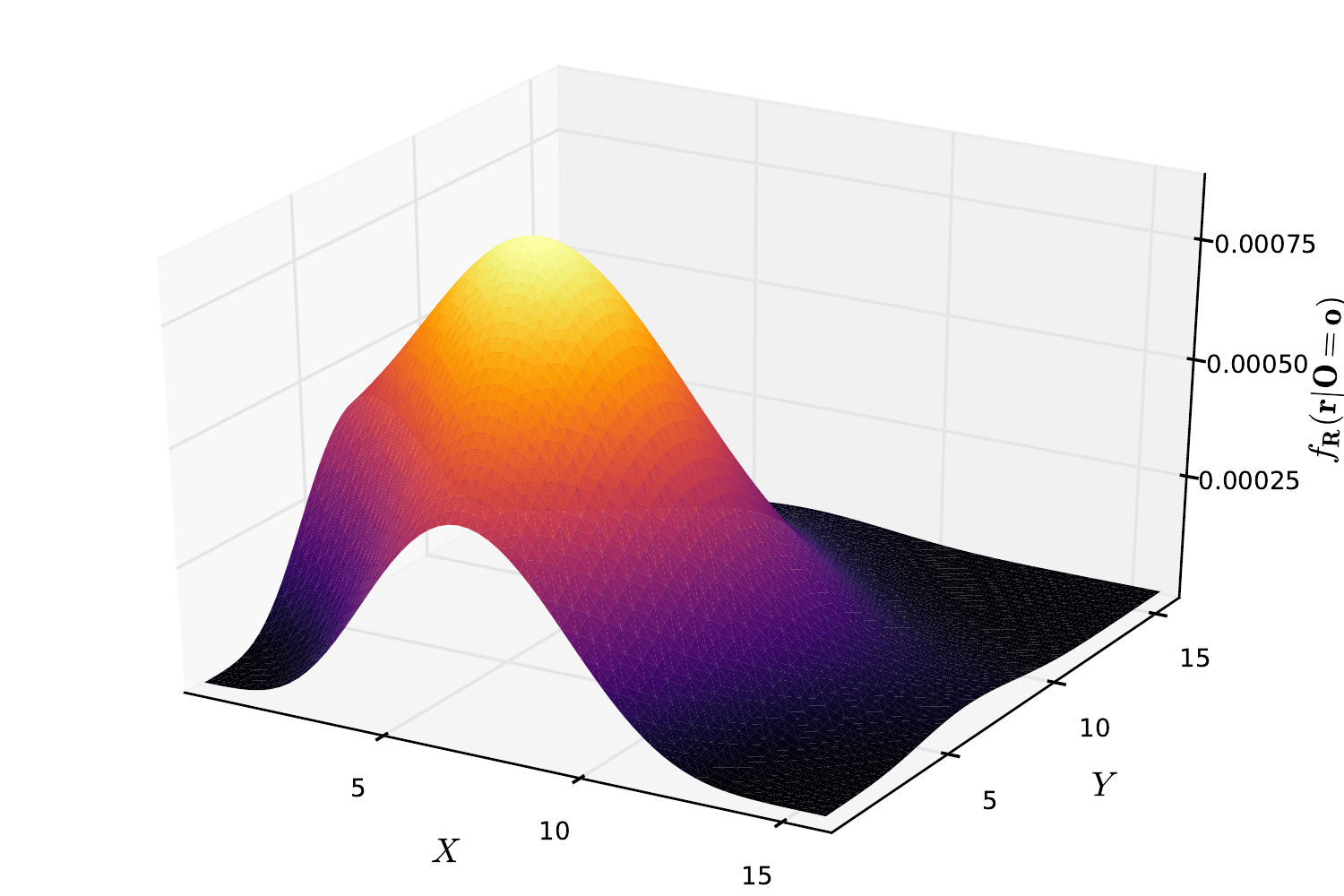}\label{fig:pdfRgO_3}}
    \hfil
    \subfloat[]{\includegraphics[width=0.49\textwidth]
    {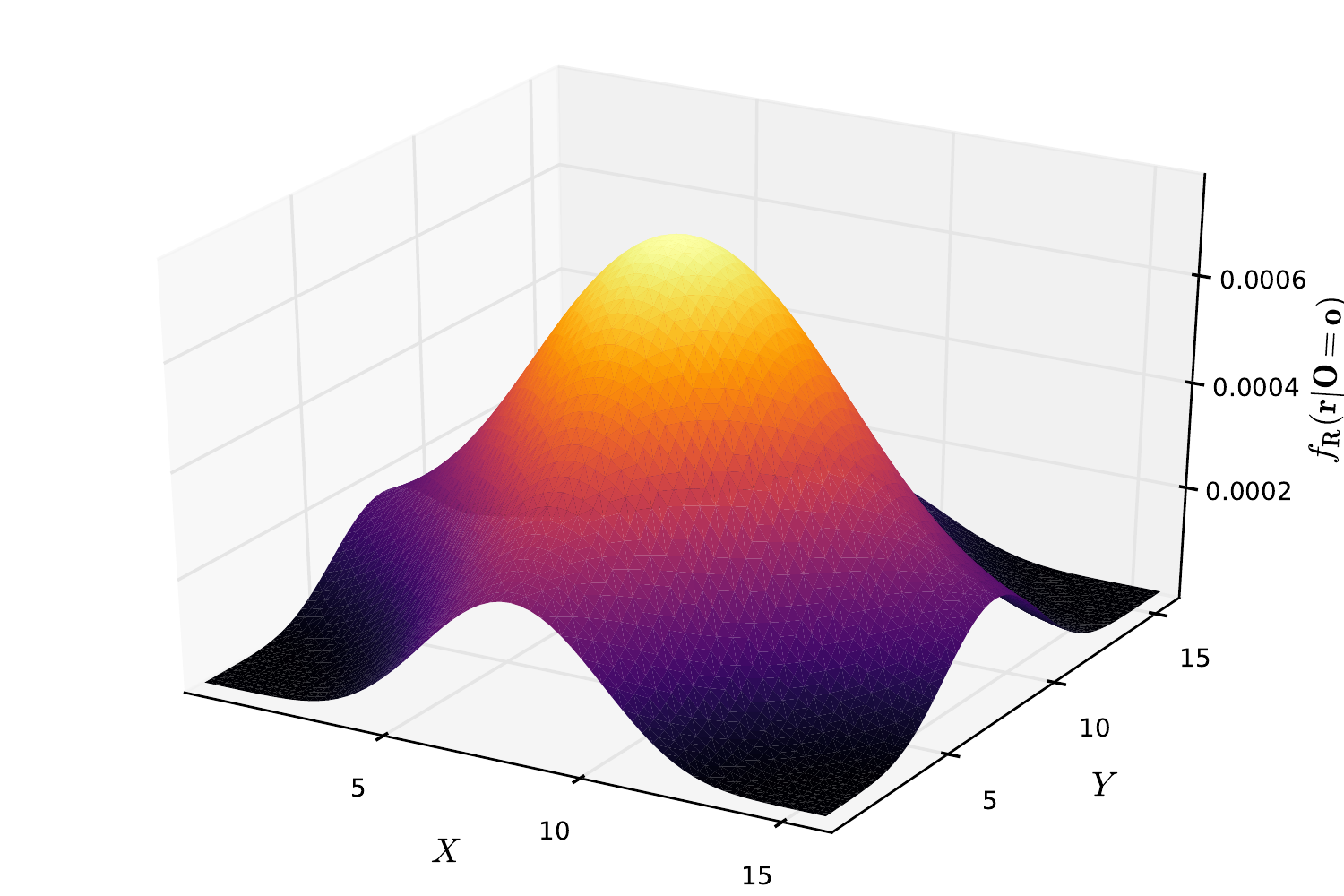}\label{fig:pdfRgO_4}}
    \caption{Illustration of the posterior distribution of $R$ using
        observations taken from a log-normal distribution. For this
        illustration, four transmitters were placed in a \SI{16}{\metre}
        $\times$ \SI{16}{\metre} area. A log-normal path loss model was used to
        to determine the signal strengths. Each subplot above shows the
        posterior distribution of $R$ constructed by the receiver upon receiving
        a different vector of observations.
    }\label{fig:pdfRgO}
\end{figure*}

Using the conditional distribution of the observed vector and the prior over $\Rv$, we
obtain the conditional distribution over the receiver locations using Bayes' rule:
\begin{align}\label{eq:rgo}
    f_{\Rv}\left( \rv \vert \Ov = \ov \right) &=
    \frac{f_{\Ov}(\ov \vert \Rv = \rv) f_{\Rv}(\rv)}
    {\int_{\rv \in \mathcal{S}} f_{\Ov}\left(
    \ov \vert \Rv = \rv \right) f_{\Rv}(\rv) \, \mathrm{d}\rv}.
\end{align}
Algorithms for localization are essentially methods that derive a location estimate from
the above posterior distribution. In fact, any localization algorithm~$A$ is a mapping from
\begin{itemize}
    \item the observation vector $\ov$
    \item the prior distribution over the location, $f_{\Rv}(\rv)$
    \item the conditional distribution over	$\ov$,
        $f_{\Ov}\left( \ov \vert \Rv = \rv \right)$
\end{itemize}
to a location estimate $\hat{\rv}$, as illustrated in
Figure~\ref{fig:algo}. A visualization\footnote{The code used to
generate the figures in this paper is available online~\cite{lczn}.} of the
posterior distribution for the popular simple path loss model with
log-normal fading is given in Figure~\ref{fig:pdfRgO}.

\subsection{Optimization based approach to Localization}

In the optimization based approach to localization that we advocate, the
starting point for estimating the receiver location is a cost function
that must be defined \emph{a priori}. In the most general terms, the
cost function is modelled as $\mathcal{C}(\rv, \tilde{\rv}, \ov)$, i.e.,
a function of the true location~$\rv$, a given proposed location
estimate~$\tilde{\rv}$, and the observation vector~$\ov$. We define the
expected cost function given an observation vector as follows:

\begin{equation}
    \E[\mathcal{C}(\rv, \tilde{\rv}, \ov)] =  \int_{\rv \in \mathcal{S}}
    \mathcal{C}(\rv, \tilde{\rv}, \ov) \; f_{\Rv}\left( \rv \vert \Ov =
        \ov \right) \mathrm{d}\rv.
\end{equation}

Given any cost function $\mathcal{C}$, the optimal location estimation
algorithm can be obtained in a unified manner by solving the following
optimization for any given observation vector to obtain the optimal estimate
$\hat{\rv}$:
\begin{equation}~\label{eq:genopt}
    \hat{\rv} =  \argmin_{\tilde{\rv}} \E[\mathcal{C}(\rv, \tilde{\rv},
    \ov)].
\end{equation}

Note that this optimization may be performed to obtain an arbitrarily
near-optimal solution by numerically computing
$\E[\mathcal{C}(\rv, \tilde{\rv}, \ov)]$ over the discretization
of a two or three dimensional search space. Given recent gains in
computing power, the optimization is feasible for
typical indoor localization problems. Moreover, the optimization
naturally lends itself to parallel execution since the computation of
the expected cost at all candidate locations are independent of each
other. Assuming uniform coverage, our solution will improve on
increasing the number of points in our search space. In practice, for
RSS localization, these points could be spaced apart on the order of
10's of centimeters.
\begin{figure}[h]
    \centering
    \includegraphics[width=0.49\textwidth]{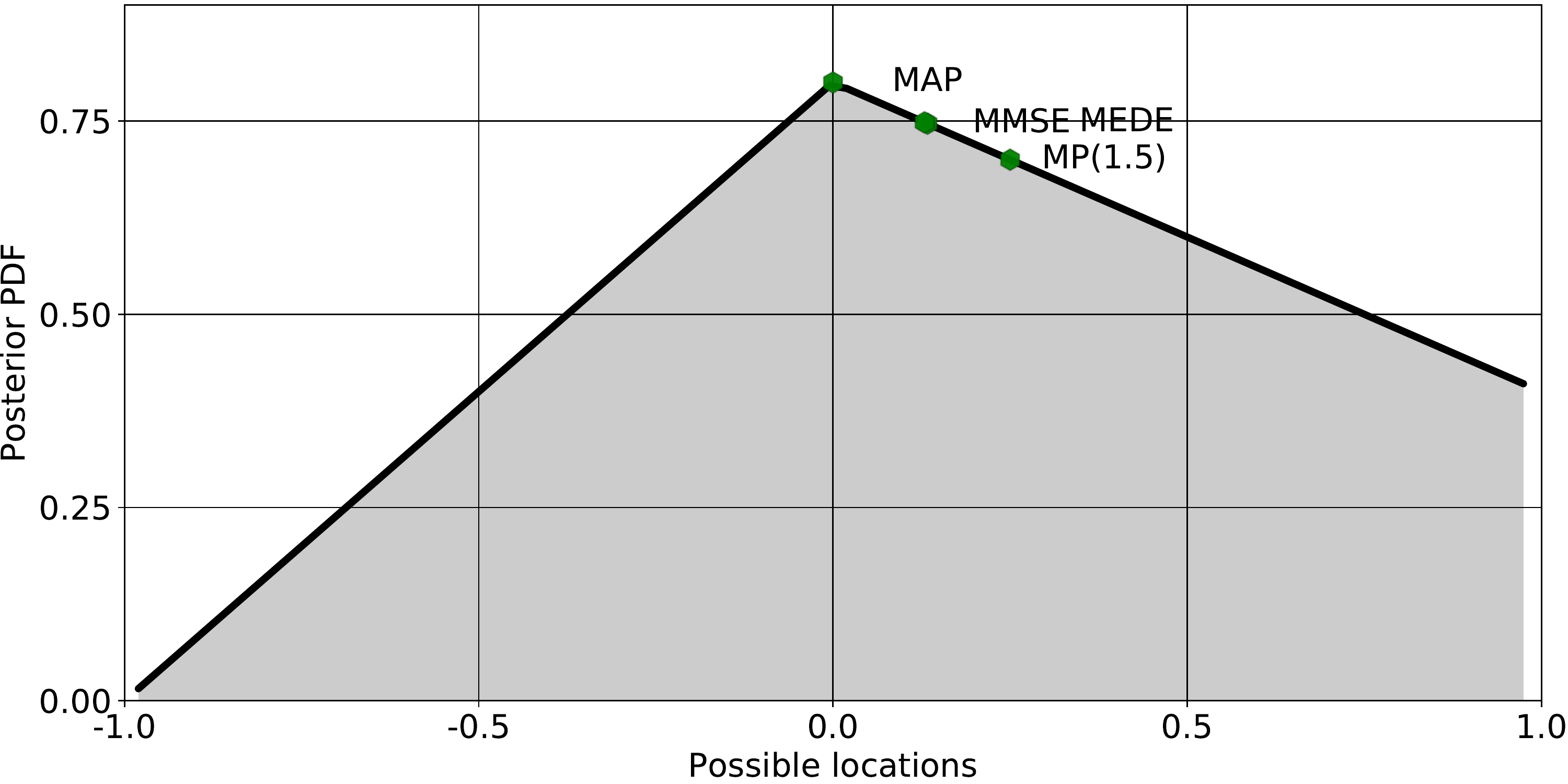}
    \caption[An illustration of the estimates returned by different
        localization algorithms.]{\leavevmode\\\begin{minipage}{\linewidth}
        An Illustration of the distribution of true location given the
        observations and the locations that correspond to different
        optimizations. In this example, we show the different estimates
        returned by the various localization algorithms for a unimodal,
        asymmetric posterior probability density function of the form
        \begin{align*}
            f(x) =
            \begin{cases}
                \frac{4}{5}(1 + x) & \quad \text{if } -1 \le x < 0 \\
                \frac{4}{5}(1 - \frac{x}{2}) & \quad \text{if } 0 \le x
                \le 1.\\
            \end{cases}
        \end{align*}
        The asymmetry of the distribution function pulls estimates other
        than MAP from the mode, with M$P(d)$ being the most
        affected. In this example, it may be shown that for $d \le 1.5$,
        the M$P(d)$ estimate is $\hat{x}_{MP(d)} = \frac{d}{6}$. Thus we
        see that the M$P(d)$ estimate moves closer to the MAP estimate with
        decreasing~$d$. This example serves to illustrate how differing
        optimization objectives can yield very different estimates for the
        same posterior distribution, thereby underlining the importance
        of deciding on an optimization objective upfront.
    \end{minipage}
    }\label{fig:pdfxgo}
\end{figure}

Existing algorithms such as MLE, MMSE, MEDE and M$P(d)$ can be recovered
in our framework using suitable choices of the cost function
$\mathcal{C}$. For instance, it is straightforward to verify that
minimizing the expected distance error yields MEDE\@. Perhaps more
interestingly, the MLE estimate can also be recovered using an
appropriate \emph{distance based} cost function. Figure~\ref{fig:pdfxgo}
provides an example of how these different optimizations can yield very
different location estimates.

The most compelling aspect of this unified optimization-based approach
to localization is its generality. Being Bayesian in nature, it can
incorporate both model and data-driven approaches to characterizing the
radio environment in a given space, and can accommodate prior
information in a natural way (as such, it is also highly compatible with
location tracking approaches that use Bayesian filtering). In addition,
the framework has cognitive properties in that it gets better over time
as more observations or inputs help improve the prior. While we present
and evaluate our framework using RSS measurements for ease of
exposition, it is not limited to such measurements. Other modalities
such as ToA, TDoA~\cite{pri00,sav01} and AoA~\cite{nic03} are easily
incorporated as well.

\section{A Unifying Framework for Evaluating Localization Algorithms}~\label{sec:errcdf}
In addition to the previously defined unifying framework, we also propose the
use of the distance \emph{error cdf} as a unified way of evaluating
localization algorithms.  For a localization algorithm, say $A$, the
$L_2$ (Euclidean) distance between the estimate ($\hat{\rv}_A$) and the true
location ($\rv$) is represented by the random variable $D_A$.
Note that
\begin{equation}
    D_A = \| \hat{\rv}_A - \rv \|_2.
\end{equation}
The CDF of $D_A$, also termed the error cdf of algorithm $A$, may be
characterized by averaging the probability that the true location lies within
a certain distance, say $d$, of our estimate, over the all possible receiver
locations. This notion is defined below.
\begin{dfn}\label{def:errcdf}
    Let $A \in \mathcal{A}$, where $\mathcal{A}$ denotes the set of all
    localization algorithms.
    Denote by $\hat{\rv}_A$ a location estimate returned by
    the algorithm $A$.
    Then, the \emph{error cdf} of $A$ is a monotonically increasing
    function $F_{A}:\mathcal{Q}\subseteq
    \mathbb{R}_{\ge 0} \to [0,1]$ such that
    \begin{equation} \label{eq:errcdf}
        F_{A}(d) = \int_{\rv \in \mathcal{S}} P\left[ D_A \le d
        \right] f_{\Rv}(\rv)\, \mathrm{d}\rv. 
    \end{equation}
\end{dfn}
Let $d^*$ be the maximum distace between any two points in
$\mathcal{S}$. Then $\mathcal{Q}$ is the closed interval $[0,d^*]$.
Using the error cdf, we may meaningfully define an ordering over the
class of localization algorithms using the concept of \emph{stochastic
dominance}.
\begin{dfn}\label{def:sd}
    Let $A_1, A_2 \in \mathcal{A}$.
    We say that $A_1$ \emph{stochastically dominates} $A_2$ if
    \begin{equation} \label{eq:sd}
        F_{A_1}(d) \ge F_{A_2}(d) \quad \forall d \in \mathcal{Q}.
    \end{equation}
\end{dfn}
\begin{dfn}
    Let $A_1, A_2 \in \mathcal{A}$.
    We say that $A_1$ \emph{strictly stochastically dominates} $A_2$ if
    in addition to equation~\eqref{eq:sd},
    there exists $d_1, d_2 \in \mathcal{Q}$ such that $d_1 < d_2$ and
    \begin{equation} \label{eq:ssd}
        F_{A_1}(d) > F_{A_2}(d) \quad \forall d \in [d_1, d_2].
    \end{equation}
\end{dfn}

\subsection{Distance Based Cost Functions}~\label{sec:costfns}
We now restrict our attention to an important class of metrics, those cost
functions that can be specified to be  monotonically increasing with respect
to the distance between the true and estimated positions.  We show that in
fact localization algorithms form a partially-ordered set with respect to this
important class of metrics. We also show that localization algorithms derived using the
optimization-based approach for these metrics lie essentially on a ``Pareto
Boundary'' of the set of all localization algorithms.
The localization cost function, formally defined below,
generalizes most metrics commonly used in the localization
literature.
\begin{dfn}~\label{def:lcf}
    Let $g:\mathcal{Q} \subseteq
    \mathbb{R}_{\ge 0} \to \mathbb{R}_{\ge 0}$ be a monotonically increasing
    function. Denote the set of all such functions by $\mathcal{G}$.
    For a localization algorithm $A$, $g(D_A)$ is the \emph{distance error
    localization cost function}. $\E\left[ g(D_A)\right]$ is the \emph{expected
    cost} of the algorithm $A$.
\end{dfn}

We use the above notion of expected cost as a metric to compare different
localization algorithms. Note that this cost function is a special case of the
more general cost function introduced in the previous section. Here we are
only interested in cost functions that depend on the distance between the true
location and our estimate.  Many localization algorithms of interest try to
optimize for some distance based cost function, either explicitly or
implicitly. We have seen already that MMSE, MEDE and M$P(d)$ have distance
based cost functions. Although perhaps not immediately apparent, MAP may also
be computed using a distance based cost function. Specifically, we may
retrieve the MAP estimate using M$P(d)$ with an adequately small radius $d$
as shown in the theorem below and also
borne out by our evaluation results.
\begin{thm}\label{thm:mapmpd}
   If the posterior distribution $f_{\Rv|\Ov}$ is continuous and the MAP
   estimate lies in the interior of $\mathcal{S}$, then for any $\delta > 0$
   there exists $\epsilon > 0$ such that,
   \begin{equation}
       \lvert P_1(d) - P_2(d) \rvert \le \delta \quad \forall \,
       0<d<\epsilon,
   \end{equation}
   where $P_1(d)$ and $P_2(d)$ give the probability of the receiver location
   being within distance $d$ of the MAP and the M$P(d)$ estimate respectively.
\end{thm}
\begin{proof}
   See Appendix.
\end{proof}

\subsection{Comparing Algorithms using Stochastic Dominance}
In this section, we explore how we may meaningfully compare algorithms
using our optimization based framework. If we are interested in a
particular cost function, then comparing two algorithms is
straightforward. Compute their expected cost and the algorithm with the
lower cost is better. However, with stochastic dominance we can deduce
something more powerful.  For any two localization algorithms $A_1$ and
$A_2$, if $A_1$ stochastically dominates $A_2$, then the expected cost
of $A_1$ is lower that that of $A_2$ for \emph{any} distance based cost
function. More formally,
\begin{thm}\label{thm:dom}
    For any two localization algorithms $A_1, A_2 \in \mathcal{A}$,
    if $A_1$ stochastically dominates $A_2$, then
    \begin{equation}\label{eq:thmdom}
        \E\left[ g(D_{A_1})\right] \le \E\left[g(D_{A_2})\right]
        \quad \forall g \in \mathcal{G}.
    \end{equation}
    If $A_1$ strictly stochastically dominates $A_2$, then
    \begin{equation}\label{eq:thmsdom}
        \E\left[ g(D_{A_1})\right] < \E\left[g(D_{A_2})\right]
        \quad \forall g \in \mathcal{G}.
    \end{equation}
\end{thm}
\begin{proof}
   See Appendix.
\end{proof}
Theorem~\ref{thm:dom} is the first step towards ranking algorithms based on
stochastic dominance. It also gives us a first glimpse of what an optimal
algorithm might look like. From Theorem~\ref{thm:dom}, an algorithm $A^*$ that
stochastically dominates every other algorithm is clearly optimal for the
entire set of distance based cost functions. However, it is not obvious that
such an algorithm need even exist. On the other hand, we can compute algorithms
that are optimal with respect to a particular cost function.
As given in the following theorem, such optimality
implies that the algorithm isn't dominated by any other algorithm. In other
words, if algorithm $A$ is optimal with respect to a distance based cost
function $g$, then $A$ is not strictly stochastically dominated by \emph{any}
other algorithm $B$.
\begin{thm}~\label{thm:aopt}
    For a localization algorithm $A \in \mathcal{A}$, if there exists a
    distance based cost function $g \in G$ such that for any other
    localization algorithm $B \in \mathcal{A}$
    \begin{align}
        \E\left[ g\left( D_{A} \right) \right] \le
        \E\left[ g\left( D_{B} \right) \right],
    \end{align}
    then for all algorithms $B \in \mathcal{A}$, there exists a distance
    $d \in Q$ such that
    \begin{align}
        F_A (d) \ge F_B (d).
    \end{align}
\end{thm}
\begin{proof}
   See Appendix.
\end{proof}
Theorems~\ref{thm:dom}~\&~\ref{thm:aopt} establish the utility of ranking algorithms
based on stochastic dominance. However, if we are given two algorithms, it is not
necessary that one should dominate the other. As Theorem~\ref{thm:nodom}
shows, if they do not conform to a stochastic dominance ordering, the
algorithms are incomparable.
\begin{thm}\label{thm:nodom}
    For any two localization algorithms $A_1$ and $A_2$, if $A_2$ does not
    stochastically dominate $A_1$ and vice versa, then there exits distance based
    cost functions $g_1, g_2 \in \mathcal{G}$ such that
    \begin{equation}\label{eq:thm2.1}
        \E\left[ g_1(D_{A_1})\right] < \E\left[g_1(D_{A_2})\right],
    \end{equation}
    and
    \begin{equation}\label{eq:thm2.2}
        \E\left[ g_2(D_{A_2})\right] < \E\left[g_2(D_{A_1})\right].
    \end{equation}
\end{thm}
\begin{proof}
   See Appendix.
\end{proof}
Theorem~\ref{thm:nodom} establishes the existence of a ``Pareto Boundary'' of
the set of all localization algorithms. Choosing an algorithm from within this
set depends on additional considerations such as its performance on specific
cost functions of interest.

\subsection{Comparison based on Upper Bound of Error
CDFs}\label{sec:upperboundcomp}
In the previous section, we focused on using stochastic dominance to rank and
compare algorithms without paying much attention to what an ideal algorithm
might look like. In this section, we explore this topic more detail.
To begin, we ask if there exists an algorithm that dominates every other algorithm?
From Theorem~\ref{thm:dom} we know that such an algorithm, if it exists, will be the best
possible algorithm for the class of distance based cost functions. Moreover the error
CDF of such an algorithm will be an upper bound on the error CDFs of all
algorithms $A \in \mathcal{A}$.
\begin{dfn}~\label{dfn:cdfBound}
    We denote the upper envelope of error CDFs for all possible algorithms
    $A \in \mathcal{A}$ by $F^*$.
\end{dfn}
We now turn our attention to formally defining the error bound $F^*$. Our definition also
provides us with a way to compute $F^*$.
Let $D_A$ represent the distance error
for algorithm $A$.
Consider the following class of M$P(d)$ cost functions. For each $d \in
\mathcal{Q}$, let
\begin{equation}~\label{eq:gddef}
    g_d\left(D\right) =
    \begin{cases}
        0 & \text{if } D \leq d \\
        1 & \text{if } D > d.
    \end{cases}
\end{equation}
Then, the value of $F^*$ at any distance $d \in \mathcal{Q}$ may be computed using the
M$P(d)$ cost
function at that distance. More formally,
\begin{dfn}
    The upper envelope of error CDFs for all possible algorithms
    $A \in \mathcal{A}$, $F^*$ is defined as
    \begin{equation} \label{eq:cdfBound}
        F^*(d) = \sup_{A \in \mathcal{A}} \left\{
        1 - \E\left[g_d(D_{A})\right]\right\}, \quad \forall d \in Q.
    \end{equation}
\end{dfn}
The upper envelope of error CDFs, $F^*$, satisfies the following properties:
\begin{enumerate}
    \item $F^*$ stochastically dominates every algorithm $A \in \mathcal{A}$,
    \item $F^*$ is monotonically increasing in $[0,d^*]$,
    \item $F^*$ is Riemann integrable over $[0,d^*]$.
\end{enumerate}
The monotonicity of $F^*$ is direct consequence of the monotonicity of
CDFs. Moreover, since $F^*$ is monotonic, it is also Riemann
integrable~\cite[p.~126]{rud76}. In general, $F^*$ may not be attainable
by any other algorithm. However, as we show below, it is achievable
under certain circumstances, which lends credence to its claim as a
useful upper bound that may be used as a basis of comparison of
localization algorithms.

Given the ideal performance characteristics of $F^*$, it is worthwhile
to investigate if it is ever attained by an algorithm. A trivial case is
when the M$P(d)$ algorithm yields the same estimate for all distances of
interest in the domain. In this particular case, MAP and M$P(d)$ are
optimal since the error CDF of the MAP or M$P(d)$ estimate traces $F^*$.
As an illustration consider a continuous symmetric unimodal posterior
distribution over a circular space with the mode located on the center
of the circle. Clearly, the MAP estimate is given by the center.
Moreover, the MP$(d)$ estimate is the same at all distances, namely the
center of the circle. Thus we immediately have that both the MAP and
M$P(d)$ estimates have attained $F^*$. An extensive discussion on the
attainability of $F^*$ can be found in Appendix~\ref{apx:attain}.

Thus we see that there exist conditions under which $F^*$ is attained by
an algorithm. Consequently, it is worthwhile to search for algorithms
that are close to this bound or even attain it under more general
settings. This leads us directly to the second method of comparing
algorithms. We identify how close the error CDFs of our algorithms get
to the upper bound $F^*$.

Consider algorithms $A,B \in \mathcal{A}$.  Intuitively, if the error
CDF of $A$ is closer to $F^*$ than that of $B$, then it seems reasonable
to expect $A$ to perform better. To make this idea precise, we need to
define our measure of closeness to $F^*$. In the following paragraph, we
propose one such measure of how close the error CDF of an algorithm $A$
is to $F^*$.  Our proposal satisfies a nice property. Namely, searching
for an algorithm that is optimal over this measure is equivalent to
searching for an algorithm that minimizes a particular distance based
cost function. Consequently, to specify the algorithm we only need to
identify this cost function.

\begin{figure*}[!t]
    \centering
    \subfloat[]{\includegraphics[width=0.49\textwidth]
    {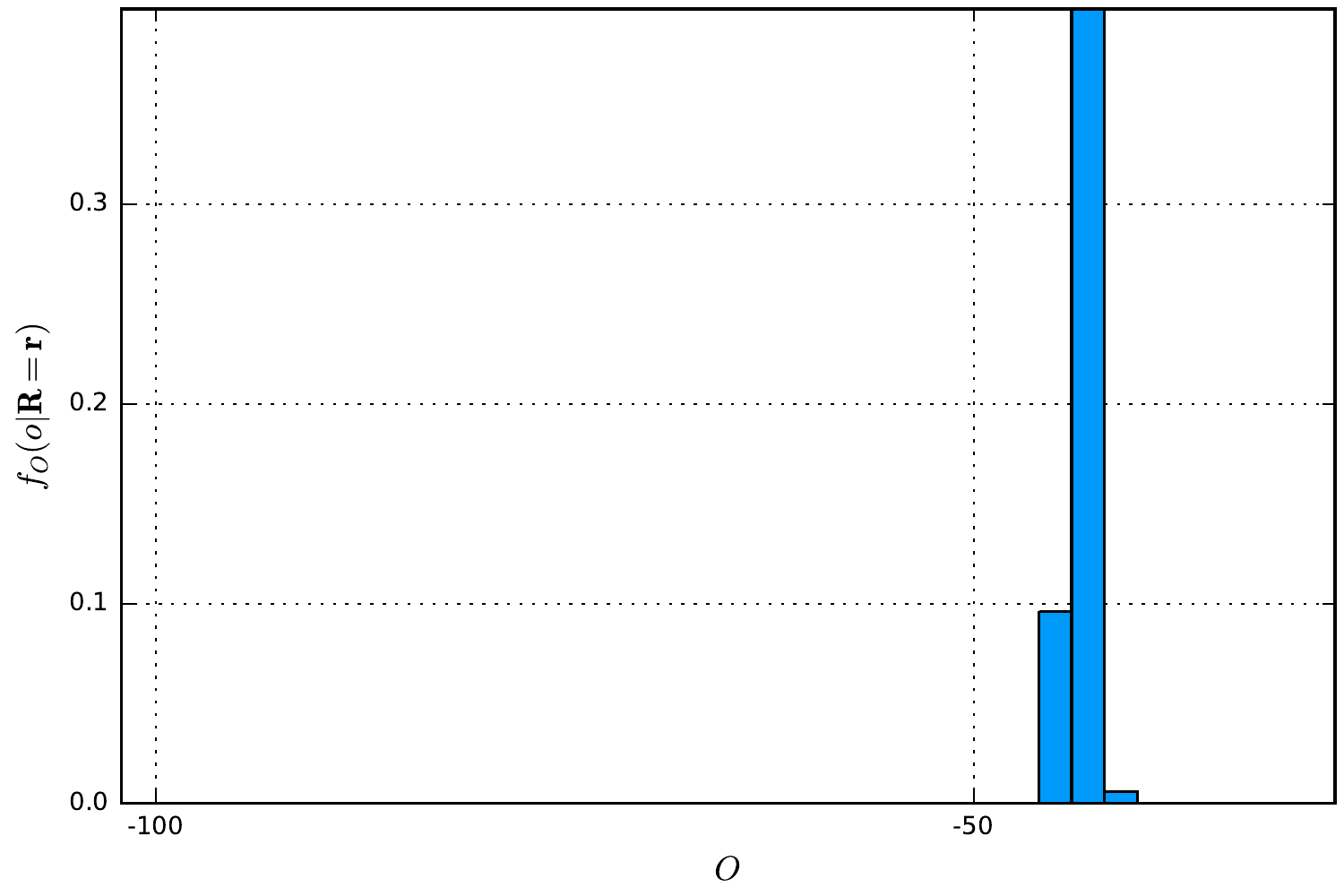}\label{fig:pdfogr_fixedr_1}}
    \hfil
    \subfloat[]{\includegraphics[width=0.49\textwidth]
    {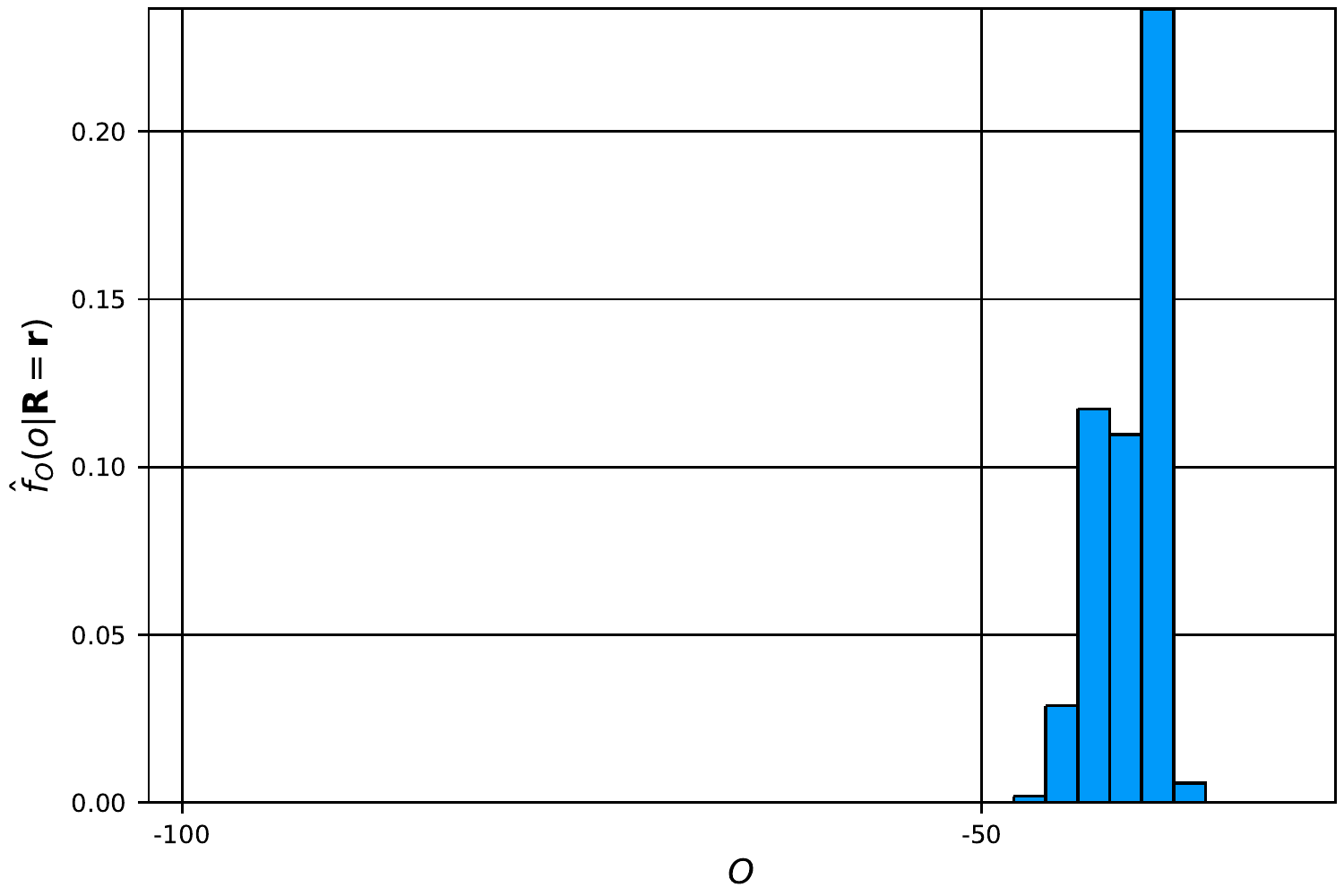}\label{fig:pdfogr_fixedr_2}}\\
    \subfloat[]{\includegraphics[width=0.49\textwidth]
    {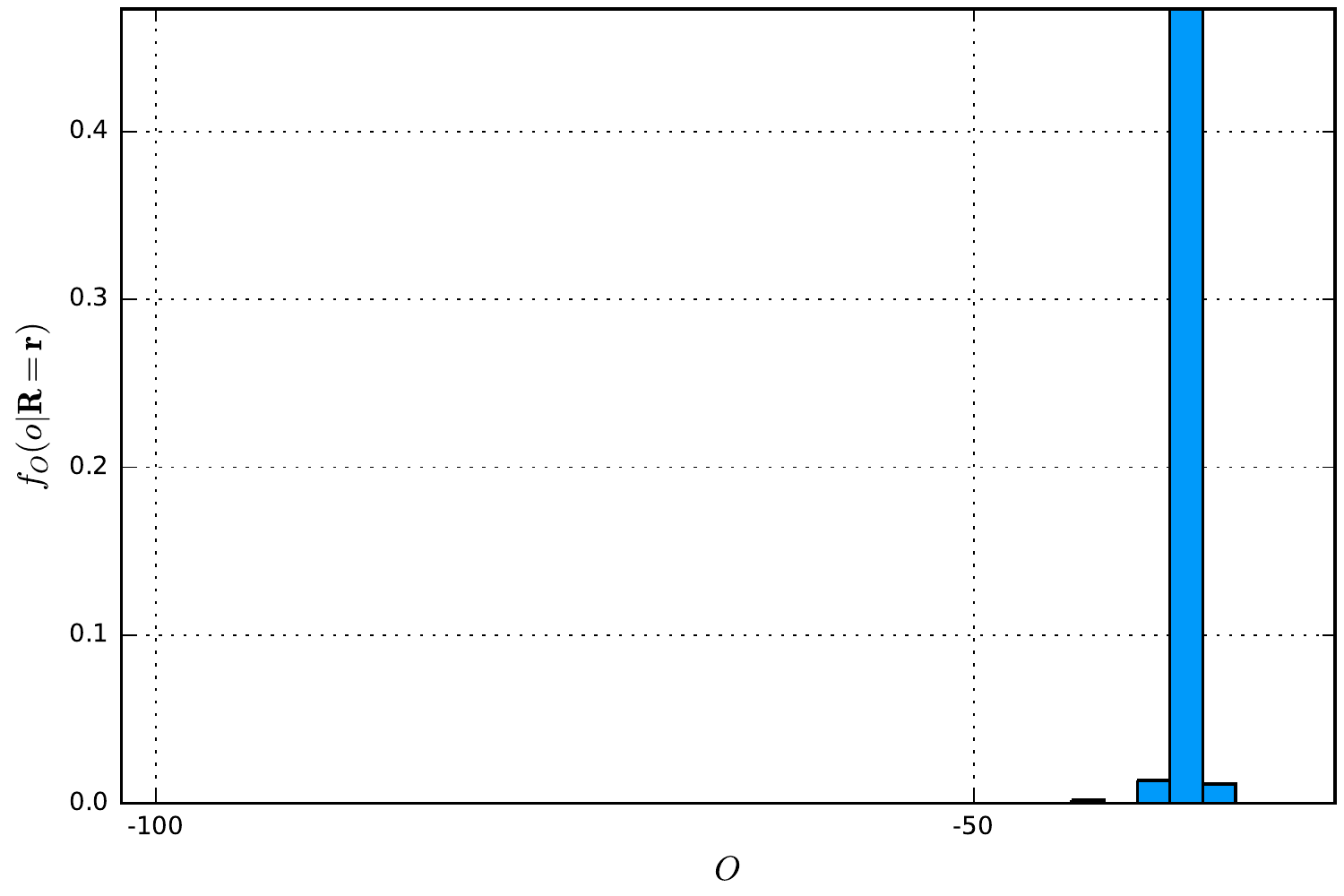}\label{fig:pdfogr_fixedr_3}}
    \hfil
    \subfloat[]{\includegraphics[width=0.49\textwidth]
    {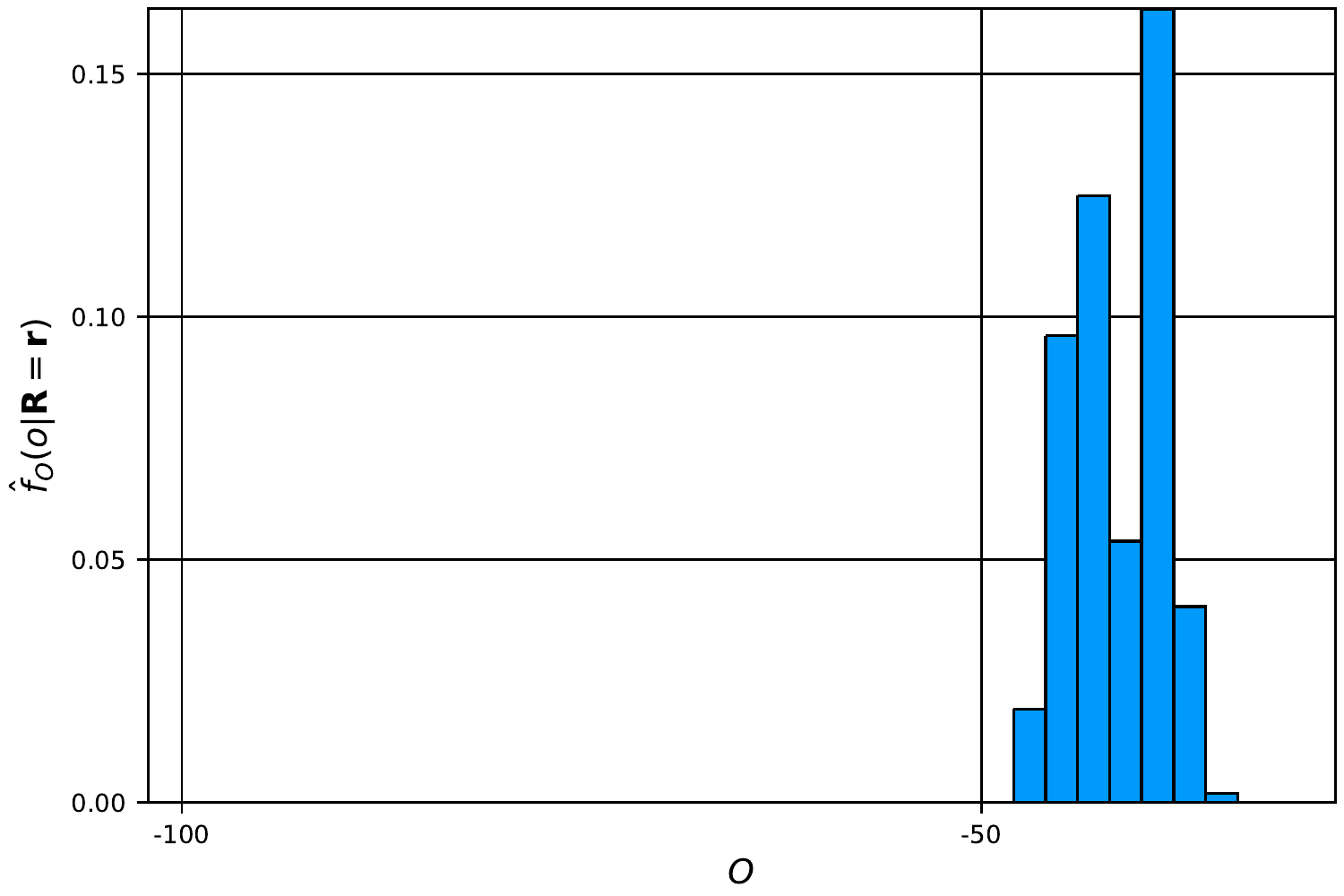}\label{fig:pdfogr_fixedr_4}}\\
    \caption{Illustration of the empirically estimated
        distribution of $O$ using signal strength measurements taken
        at different locations. Each subplot refers to the distribution
        of signal strength from a unique access point. The left subplots
        refer to measurements taken at night, while the right subplots
        refer to measurements taken during daytime.
        Our framework makes better use of the higher variance data.
    }\label{fig:pdfogr_fixedr}
\end{figure*}

\begin{dfn}
The area between the error CDF of algorithm $A$ and the upper envelope of
error CDFs is given by
\begin{equation} \label{eq:msure}
    \Theta_A = \int_0^{d^*} \left( F^*(x) - F_A(x) \right) \mathrm{d}x.
\end{equation}
\end{dfn}
The intuition behind our measure is can be summarized easily. We seek to find
an algorithm $A$ that minimizes the ``area enclosed'' by $F^*$ and the error
CDF of $A$.  Note that $\Theta_A \ge 0$ for all $A \in \mathcal{A}$. In general, it
is not clear if every measure of closeness between $F^*$ and $F_A$ will yield
a cost function for us to minimize. However, if we do find such a cost
function, then we have the advantage of not needing to explicit know $F^*$ in
the execution of our algorithm. This is the case for $\Theta_A$ as proved in the
theorem below.

\begin{thm}~\label{thm:area}
    The algorithm that minimizes the area between its error CDF and the upper
    envelope of error CDFs for all possible algorithms is the MEDE algorithm.
\end{thm}
\begin{proof}
   See Appendix.
\end{proof}

As consequence of Theorem~\ref{thm:area}, we note that if $F^*$ is
attainable by any localization algorithm, then it is attained by MEDE\@.
This in turn yields a simple test for ruling out the existence of an
algorithm that attains $F^*$. On plotting the error CDF plots of
different algorithms, if we find an algorithm that is not dominated by
MEDE, then we may conclude that $F^*$ is unattainable. Thus it is
relatively easy to identify cases where there is a gap between $F^*$
and MEDE\@. However, the issue of confirming that MEDE has
attained $F^*$ is more difficult as it involves a search over the set of
all algorithms.

In summary, the utility of $F^*$ lies in its ability to pin point the
strengths and weaknesses of a proposed algorithm. As we have seen, some
algorithms such as M$P(d)$ is designed to do well at specific distances
while others such as MEDE aims for satisfactory performance at all
distances. Other algorithms lie somewhere in between. Consequently,
choosing one algorithm over the other depends on the needs of the
application utilizing the localization algorithm. Therein lies the
strength of our proposed framework. It allows us to effectively reason
about the applicability of an algorithm for the use case at hand.

\section{Evaluation}~\label{sec:evaluation}
\begin{figure*}[!t]
    \centering
    \subfloat[]{\includegraphics[width=0.49\textwidth]
    {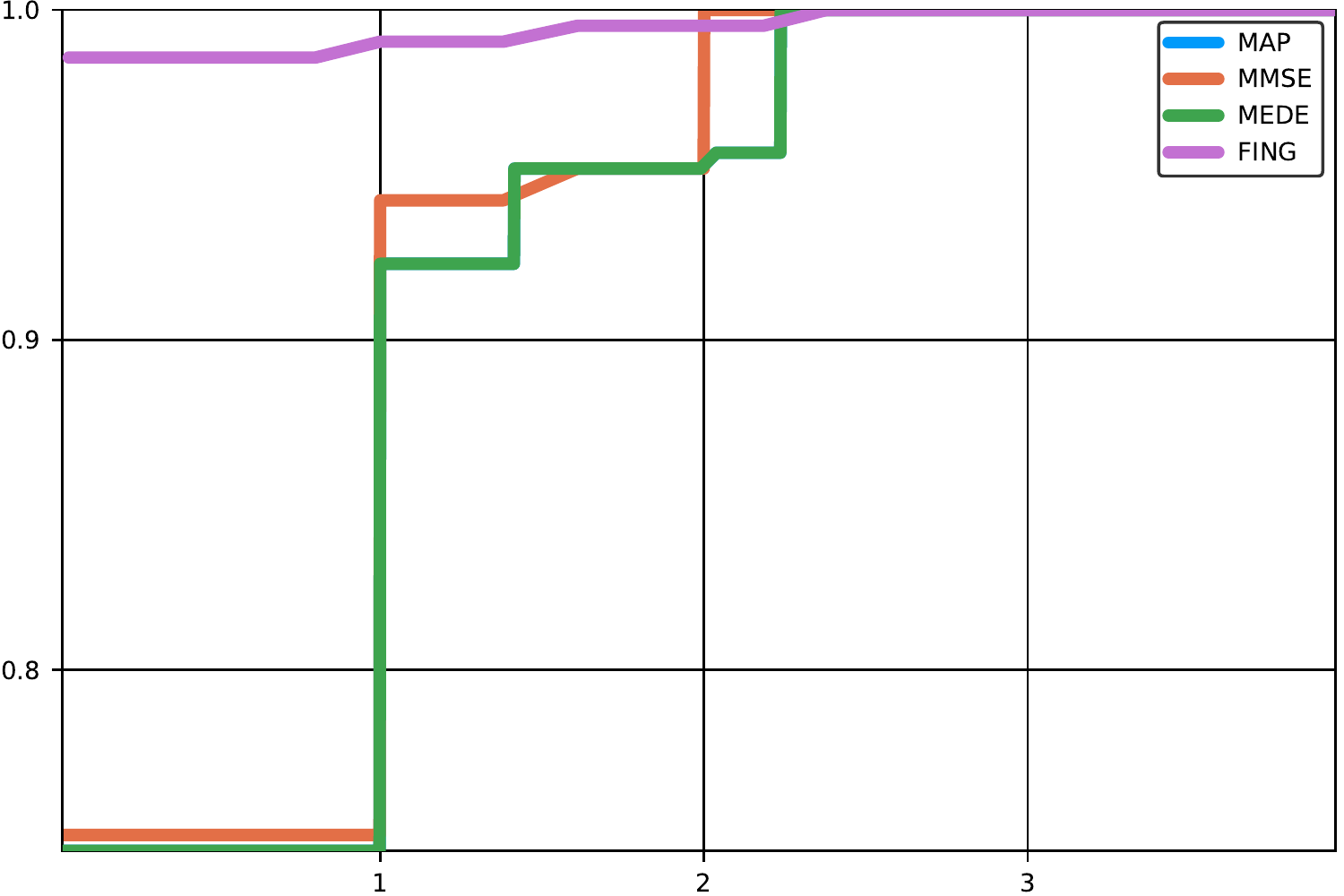}\label{fig:ecdf_lowvar_1}}
    \hfil
    \subfloat[]{\includegraphics[width=0.49\textwidth]
    {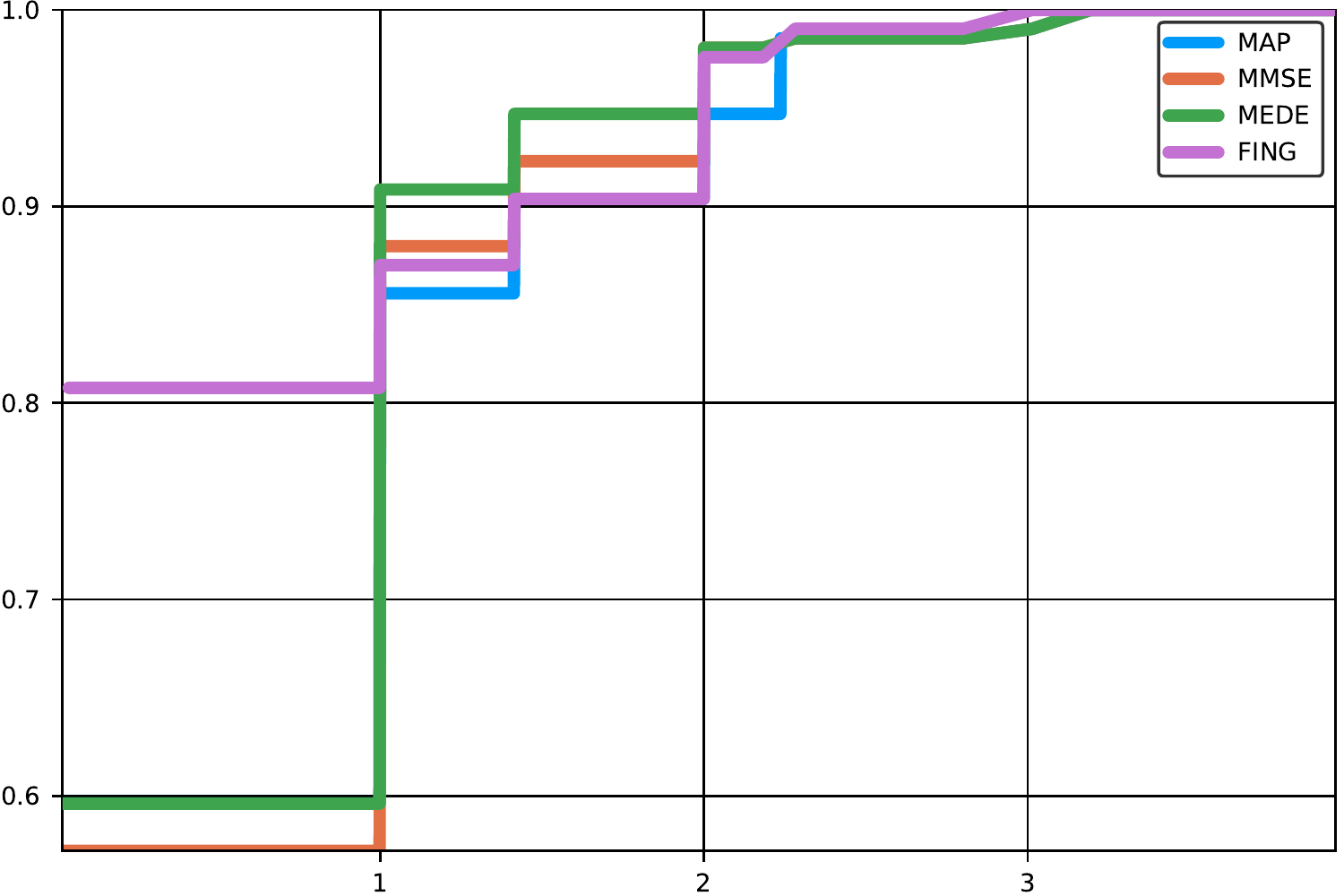}\label{fig:ecdf_lowvar_2}}\\
    \subfloat[]{\includegraphics[width=0.49\textwidth]
    {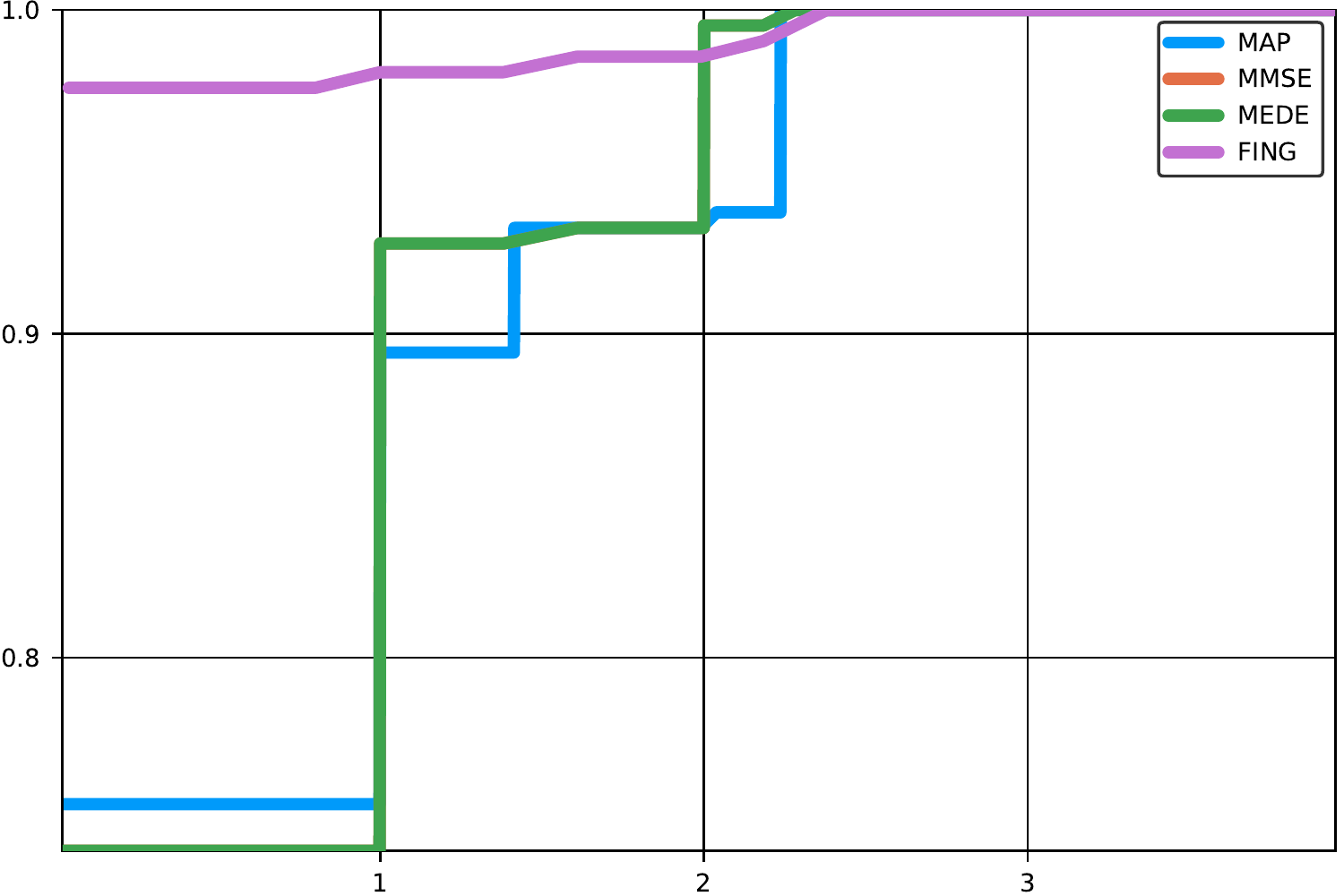}\label{fig:ecdf_lowvar_1}}
    \hfil
    \subfloat[]{\includegraphics[width=0.49\textwidth]
    {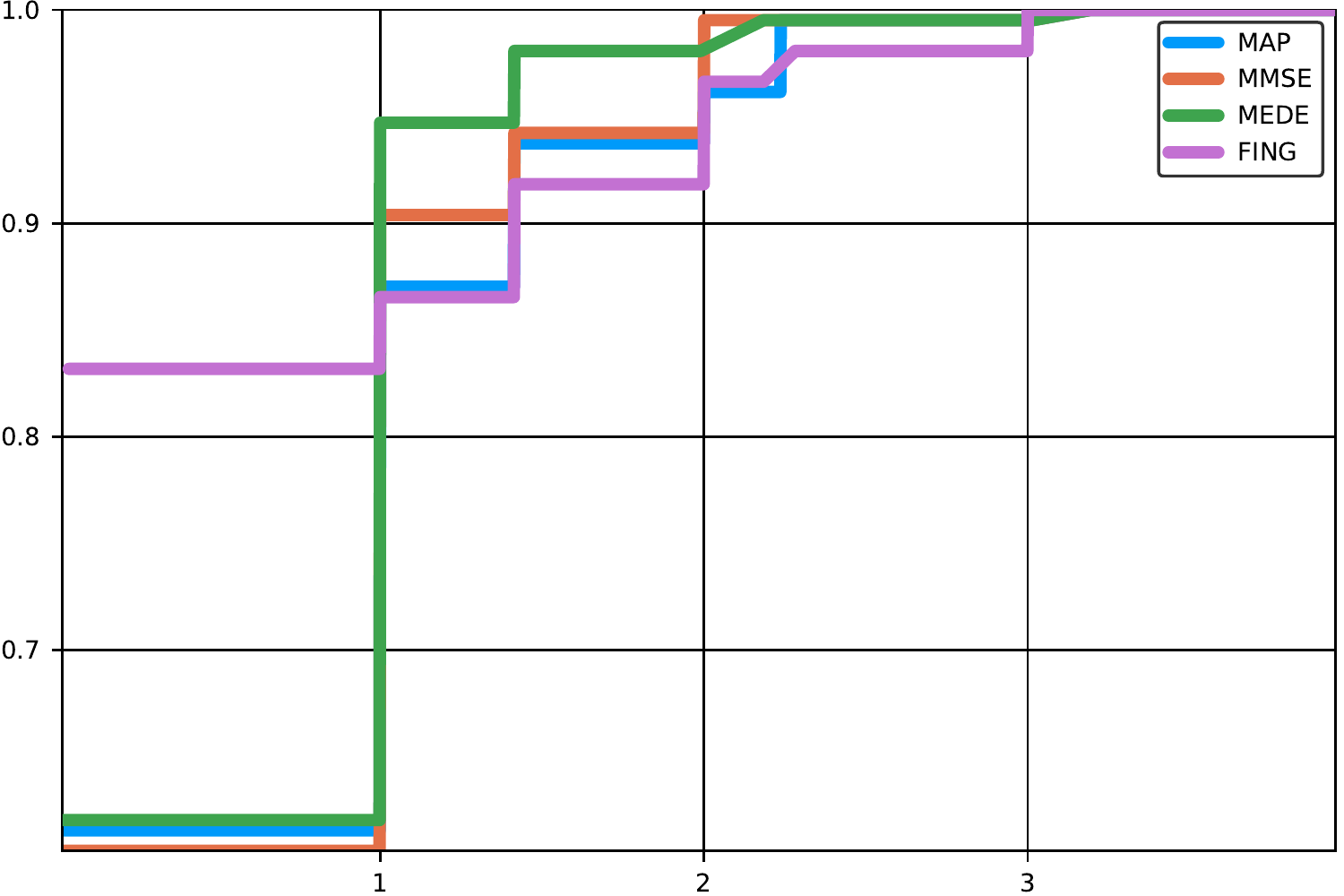}\label{fig:ecdf_lowvar_2}}\\
    \caption{Comparison of error CDFs for different localization
        algorithms. Traditional fingerprinting based on matching the
        mean signal strengths is indicated by the moniker `FING'. Each
        subplot indicates the algorithm performance for a different set
        of test data. The left subplots refer to the case when the
        signal strengths are tightly clustered around the mean, while
        the right subplots refer to the measurements with more variance.
    }\label{fig:ecdf_lowvar}
\end{figure*}

We evaluate the proposed framework using simulations, traces and real
world experiments. In Section~\ref{ssec:finger}, we provide an
illustration of how fingerprinting methods fit in to the framework
presented in Section~\ref{sec:framework}, using real-world data
collected from an indoor office environment. We show that while many
implementations of fingerprinting use only the mean signal strength from
each transmitter, we are able to better utilize the collected data by
building an empirical distribution of the received observations.

We also evaluate the performance of the MLE, M$P(d)$, MMSE and MEDE
using simulations as well as using traces~\cite{bau09}. In both cases
the signal propagation was modelled using a simplified path loss model
with log-normal shadowing~\cite{mol10}. We also assume that the prior
distribution ($f_{\Rv}$) is uniform over $\mathcal{S}$.

\subsection{Fingerprinting Methods}\label{ssec:finger}
Model-based methods assume that the distribution of observations given a
receiver location is known. In contrast, fingerprinting methods avoid
the need to model the distribution of observations by noting that one
only needs to identify the change in distribution of observations from
one location to another. Most implementations simplify matters even
further by \emph{assuming} that mean of the observations is distinct
across different locations in our space of interest. The estimated mean
is thus said to `fingerprint' the location.

This approach works well only in cases when the distribution of signal
strengths is mostly concentrated around the mean. In this case, the
approach of using only the mean signal strength amounts to approximating
the signal strength distribution with a normal distribution centered at
the estimated mean signal strength and variance approaching zero.
However, if the distribution has significant variance this approach is
likely to fail. Indeed, in the regime of significantly varying signal
strengths, keeping only the mean amounts to throwing away much of the
information that one has already taken pains to collect.

As already indicated in Section~\ref{sec:framework}, we make better use
of the collected data by empirically constructing the distribution of
observations $f_{\Ov}\left( \ov \vert \Rv = \rv \right)$. This
formulation allows us to the use the same algorithms as in the
model-based approach, as the only difference here is in the construction
of $f_{\Ov}\left( \ov \vert \Rv = \rv \right)$. This is in contrast to
many existing implementations where one resorts to heuristics such as
clustering. Indeed, under mild assumptions it is well known that the
empirical distribution converges with probability one to true
distribution~\cite{tuc59} which gives our approach the nice property
that it can always do better given more data.

As a proof-of-concept, we compare the performance of our approach with
that of traditional fingerprinting methods in two different settings.
The data was collected from a $\SI{4}{\metre} \times \SI{2}{\metre}$
space inside an office environment. The space was divided into eight
$\SI{1}{\metre} \times \SI{1}{\metre}$ squares and signal strength
samples were collected from the center of each square. Two hundred and
fifty signal strength readings were collected for the ten strongest
access points detected using the WiFi card on a laptop running Linux.
The beacon interval for each access point was approximately
$\SI{100}{\milli\second}$. The signal strength measurements were taken
$\SI{400}{\milli\second}$ apart. Two sets of data were collected, one at
night time and the other during the day. The measurements taken at night
show that the observed signal strengths are highly
concentrated around the mean, as can be seen from the left subplots of
Figure~\ref{fig:pdfogr_fixedr}. The measurements taken during daytime
show slightly more variability as can be seen in the right subplots of
Figure~\ref{fig:pdfogr_fixedr}.

Ten percent of the collected data is randomly chosen for evaluating
algorithm performance. The remaining data
is used to construct the empirical distribution
$\hat{f}_{\Ov}\left( \ov \vert \Rv = \rv \right)$ from which the MMSE,
MAP and MEDE estimates are derived. It is also used to compute the mean
signal strength vector or fingerprint for each location. For the
algorithm denoted as `FING' in Figure~\ref{fig:ecdf_lowvar}, the fingerprint
closest to the test observation vector (in terms of Euclidean distance)
is used to predict the location.

From the performance results given in Figure~\ref{fig:ecdf_lowvar}, we
see that, as expected, the traditional fingerprinting approach
works very well when the variability in the signal strength data is low.
On the other hand, even with slight variability in the data, the
estimates derived using our Bayesian framework outperforms traditional
fingerprinting; thereby demonstrating how we may make better use of
collected observation data using our framework.

\subsubsection{Cognitive Self-Improvement}
\begin{figure}[t]
    \centering
    \includegraphics[width=0.49\textwidth]{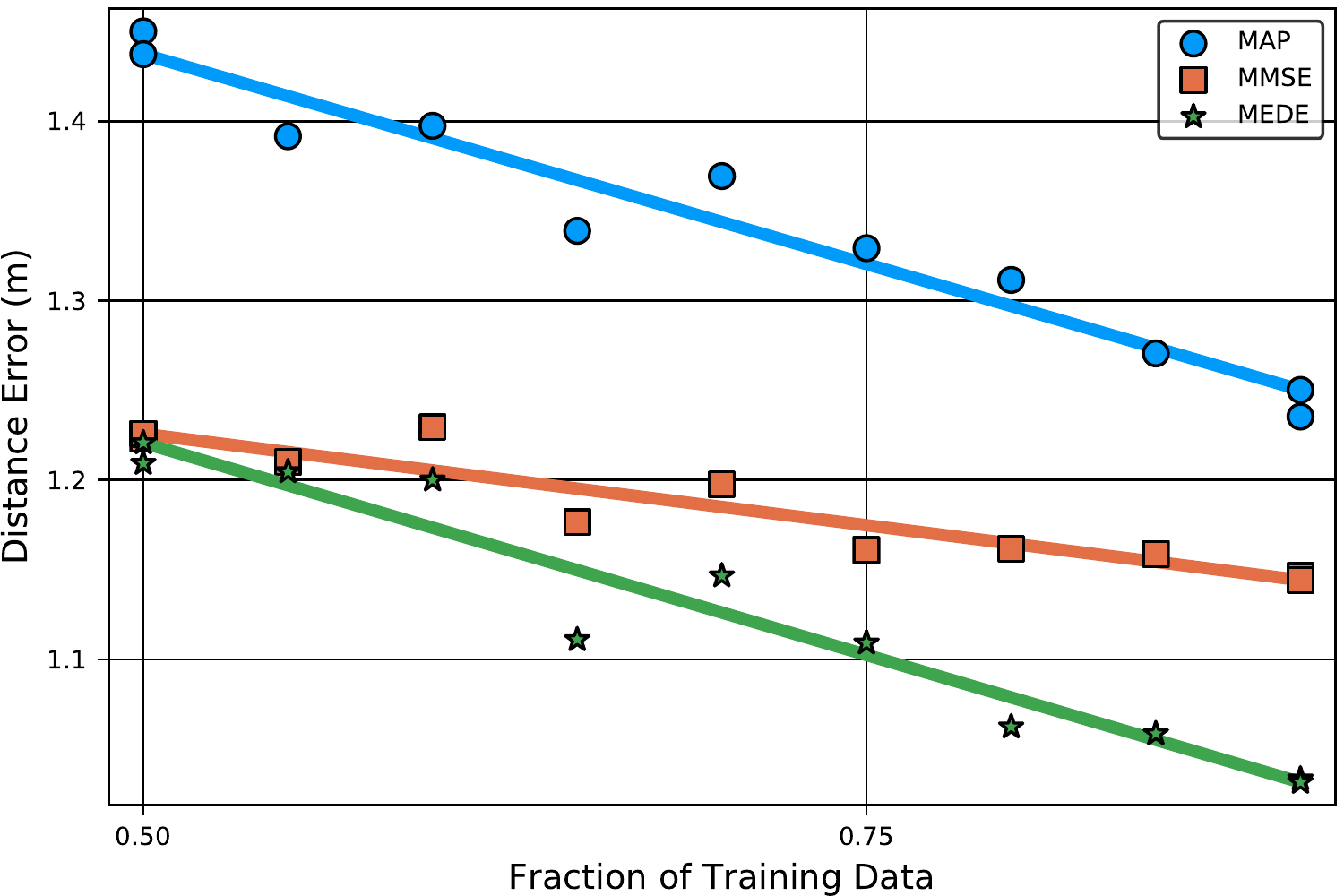}
    \caption{Variation of distance error with the size of the training
    data set. For each fraction of the original data set,
    we compute the distance error for 100 random choices of the data
    points. The regression line for each algorithm is also shown. The
    results indicate that our empirical estimates of the prior improves
    with increasing training data, which results in better algorithm
    performance.
    }\label{fig:cogimp}
\end{figure}

To illustrate the cognitive self-improving property of our approach, in
that it performs better given more observations over time, we
investigate the variation of distance error with increasing size of the
training data set. The data set with more variability was chosen for the
purposes of this illustration.
For each fraction of the original data set,
we compute the distance error for 100 random choices of the data
points, for MAP, MEDE and MMSE algorithms.
Figure~\ref{fig:cogimp} shows how the average of these distance errors
varies on increasing the size of the training data set. As can be seen
from Figure~\ref{fig:cogimp}, with increasing data we are able to better
estimate the empirical distribution
$\hat{f}_{\Ov}\left( \ov \vert \Rv = \rv \right)$ from which the MMSE,
MAP and MEDE estimates are derived, thereby resulting in better
performance.

\subsection{Simulation Model}
Say $\left\{ \lv_1, \lv_2, \dots, \lv_N \right\} \: \left( N>2 \right)$
are the known positions of $\left( N \right)$ wireless transmitters. We
assume each transmitter is located on a planar surface given by
$\mathcal{S} = [0, l] \times [0, b]$ where $l,b \; \in \mathbb{R}_{>
\zerov}$. The locations of the transmitters are given by the two
dimensional vector $\lv_i = (x_i, y_i) \in \mathcal{S} \; \forall i \in
\left\{ 1,2,\dots,N \right\}$. We wish to estimate the receiver
locations, given by the vector  $\rv = \left( x, y \right)$, from the
received signal strengths. For a given transmitter-receiver pair, say
$i$, the relationship between the received signal power ($P_r^i$) and
the transmitted signal power ($P_t^i$) may be modelled by the simplified
path loss model:
\begin{equation} \label{eq:pl}
    P_r^i = P_t^i K {\left[ \frac{d_0}{d_i} \right]}^\eta W_i,
\end{equation}
where the distance between the receiver and the $i^{th}$
transmitter is given by
\begin{equation} \label{eq:dist}
    d_i\left( \rv \right) = \sqrt{{\left( x-x_{i} \right)}^2 +  {\left(
    y-y_{i} \right)}^2 },
\end{equation}
and $W_i$ represents our noise that is log-normally
distributed with zero mean and variance $\sigma^2$.
Taking logarithm on both sides of~\eqref{eq:pl}, we get
\begin{equation} \label{eq:pldb}
    P_r^i|_{\si{\dbm}} = P_t^i |_{\si{\dbm}} + K|_{\si{\db}} - 10\eta\log_{10}
    \left[\frac{d_i}{d_0} \right] + W_i|_{\si{\db}},
\end{equation}
where $K$ is a constant given by the gains of the receiver and transmit
antennas and possibly the frequency of transmission. $d_0$ is a
reference distance, taken to be $1\si{\meter}$. In this
setting, our estimation problem may be restated as follows. We are given
measurements of the receiver signal strengths $\left\{ P_r^1,
P_r^2,\dots,P_r^N \right\}$ from which we are to estimate the receiver
location $\rv$. Thus, our observation vector $\Ov$ may be written as
\begin{align} \label{eq:pln}
    O_i
    &= P_r^i|_{\si{\dbm}} - P_t^i |_{\si{\dbm}} -
      K|_{\si{\db}} \\
    &= W_i|_{\si{\db}} - 10\eta\log_{10}\left[
      \frac{d_i}{d_0} \right],
\end{align}
for all $i \in \{1,\dots,N\}$.
In other words, the distribution of each observation is given by
\begin{equation} \label{eq:simpdfogr}
    O_i \sim \mathcal{N}\left( -10\eta\ln \left[ d_i\left( \rv \right)
    \right], \sigma^2 \right).
\end{equation}
Finally, the distribution of the observation vector $f_{\Ov}(\ov \vert
\Rv = \rv)$ can be obtained from the above by taking the product of all
the individual observation pdfs.

\subsection{Simulation and Trace Results}

\begin{table}[h]
\centering
\renewcommand{\arraystretch}{1.5}
\caption{Normalized Performance Results}
\label{tab:results}
\begin{tabular}{@{} lrrrrr @{}} \toprule
    \multicolumn{6}{c}{Simulations} \\
    \cmidrule{2-6}
    & Likelihood & $P(\epsilon)$ & $P(d)$ & MSE & EDE \\
    \midrule
    MLE            & 1.0000 & 0.9222 & 0.8994 & 1.4359 & 1.1240 \\
    M$P(\epsilon)$ & 0.9808 & 1.0000 & 0.9165 & 1.3172 & 1.0997 \\
    M$P(d)$        & 0.6963 & 0.7573 & 1.0000 & 1.2860 & 1.0857 \\
    MMSE           & 0.6806 & 0.6980 & 0.8737 & 1.0000 & 1.0643 \\
    MEDE           & 0.7247 & 0.7455 & 0.9080 & 1.1272 & 1.0000 \\
    \midrule
    \multicolumn{6}{c}{Traces} \\
    \cmidrule{2-6}
    & Likelihood & $P(\epsilon)$ & $P(d)$ & MSE & EDE \\
    \midrule
    MLE            & 1.0000 & 0.9989 & 0.9865 & 1.1583 & 1.1808 \\
    M$P(\epsilon)$ & 0.9976 & 1.0000 & 0.9881 & 1.1522 & 1.1785 \\
    M$P(d)$        & 0.8213 & 0.8596 & 1.0000 & 1.2605 & 1.3760 \\
    MMSE           & 0.9013 & 0.9171 & 0.9797 & 1.0000 & 1.1101 \\
    MEDE           & 0.8529 & 0.8685 & 0.9517 & 1.1569 & 1.0000 \\
\bottomrule
\end{tabular}
\end{table}

The parameters for the simulation were chosen to be identical as that of
the traces. The dimensions of the area of interest ($\mathcal{S}$) was
$\SI{50}{\metre} \times \SI{70}{\metre}$. Sixteen transmitters were
chosen randomly and $100$ RSSI readings were taken for each transmitter
at $5$ distinct receiver locations. The transmit power was kept constant
at $\SI{16}{\dbm}$. The estimated model parameters were a path loss of
$K = \SI{39.13}{\db}$ at reference distance $d_0 = \SI{1}{\metre}$,
fading deviation $\sigma = 16.16$ and path loss exponent $\eta = 3.93$.
We used two distances for the M$P(d)$ algorithm: (i) $\epsilon =
\SI{0.5}{\metre}$ was relatively small while (ii) $d = \SI{3}{\metre}$
covers a more sizable area. The normalized performance results are
presented in Table~\ref{tab:results}.  Each row evaluates the
performance of the indicated algorithm across different metrics, while
each column demonstrates how different algorithms perform under the
given metric. The performance value for each metric is normalized by the
performance of the best algorithm for that metric. Thus the fact that
each algorithm performs best in the metric that it is optimized for, is
reflected in the occurrence of ones as the diagonal entries in the
table.

As the algorithms presented here are each optimal for a specific cost
function, the theory predicts
that none of them are strictly stochastically dominated by any other
algorithm. The results confirm this theoretical prediction.
Also worthy of note is the similarity in the performance of
MLE and M$P(\epsilon)$, which is in line with what we expect from
Theorem~\ref{thm:mapmpd}.
\begin{figure}[ht]
    \centering
    \includegraphics[scale=0.18]{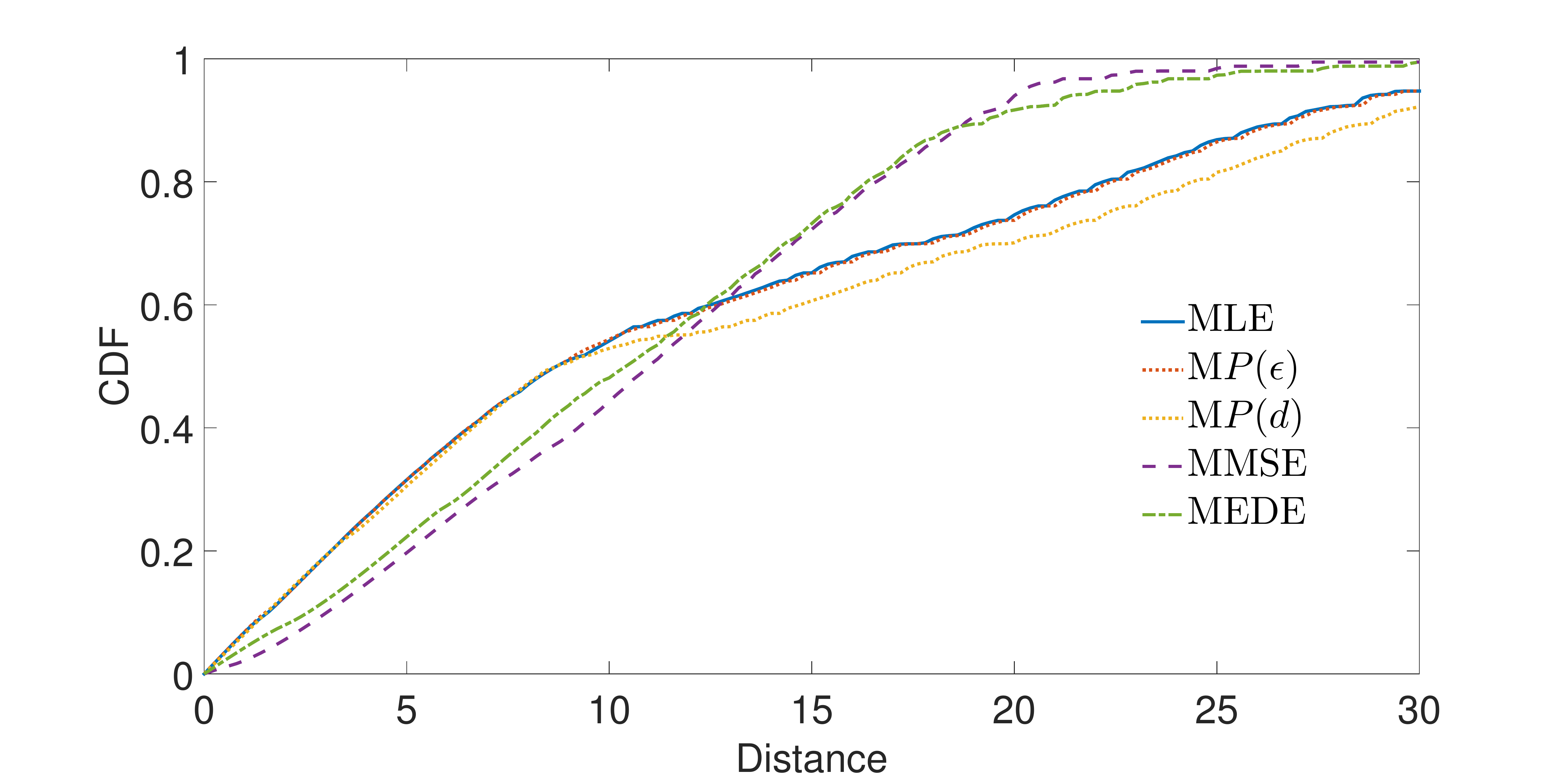}
    \caption{The error CDFs of MLE, M$P(\epsilon)$, MMSE and MEDE\@. For
        this illustration, nine transmitters were placed evenly in a
        line and log-normal fading was assumed. Note that none of the
        given algorithms are strictly stochastically dominated. In
    addition, MLE and M$P(\epsilon)$ perform identically as
expected.}\label{fig:1Dcdfall}
\end{figure}

In Figure~\ref{fig:1Dcdfall} we show the error CDFs of MLE,
M$P(\epsilon)$, MMSE and MEDE for a linear topology. As expected, no
algorithm is dominated by another and choosing $\epsilon$ to be small
results in the performance of M$P(\epsilon)$ being near identical to
that of MLE\@.  Intuitively, M$P(d)$ tries to identify the region with
given radius that `captures' most of the \emph{a posteriori} pdf
$f_{\Rv}(\rv \vert \Ov = \ov)$. Consequently for a sufficiently small
radius $d$, M$P(d)$ will return a region that contains the MLE estimate.
As a result, we are justified in thinking of the M$P(d)$ algorithm as a
generalization of MLE (or MAP, in case our prior is non-uniform).  In
practice, the value of $d$ to be used will be dictated by the needs of
the application that makes use of the localization service.

\section{Conclusion}~\label{sec:conclusion}
We have introduced an optimization based approach to localization using
a general unified framework that also allows for a fair and consistent
comparison of algorithms across various metrics of interest. We have
demonstrated how this framework may be used to derive localization
algorithms, including fingerprinting methods. We have shown the
existence of a partial ordering over the set of localization algorithms
using the concept of stochastic dominance, showing further that the
optimality of an algorithm over a particular distance based cost
function implies that the algorithm is not stochastically dominated by
another. We have identified key properties of an ``ideal'' localization
algorithm whose performance corresponds to the upper bound on error
CDFs, and highlighted how we may compare different algorithms relative
to this performance. Specifically, we have shown that MEDE minimizes the
area between its own error CDF and the performance of such an ideal
algorithm. We believe that the framework presented here goes a long way
towards unifying the localization literature. The optimization based
approach places the localization algorithm desideratum at the forefront,
where we believe it belongs.

\appendices%

\section{Proof of Theorem~\lowercase{\ref{thm:mapmpd}}}
For ease of notation, denote the posterior distribution $f_{\Rv \vert
\Ov}$ by $f$. Define the open ball of radius $d>0$
around $\rv$ as
\begin{equation}
    B_d(\rv) = \left\{\xv \in \mathcal{S} \mid \|\xv - \rv\|_2 <
    d\right\}.
\end{equation}
We denote the interior of $\mathcal{S}$ by $\interior{S}$. Let $\rv_1
\in \interior{S}$ be the MAP estimate. Pick any $\delta > 0$.
Define
\begin{equation}
    \delta' = \frac{\delta}{\int_{\rv \in \mathcal{S}} \mathrm{d}\rv}.
\end{equation}
From the continuity of $f$ there exists $\epsilon > 0$ such that
\begin{equation} \label{eq:f(map)}
    f(\rv_1) - f(\rv) < \delta' \quad \forall \,
    \|\rv_1 - \rv\|_2 < \epsilon.
\end{equation}
Consequently for any $0<d<\epsilon$ we have,
\begin{equation}
    \int_{\rv \in B_{d}(\rv_1)}\left( f(\rv_1)-\delta' \right)
    \, \mathrm{d}\rv \le 
    \int_{\rv \in B_{d}(\rv_1)} f(\rv) \, \mathrm{d}\rv.
\end{equation}
Let $\rv_2$ be the M$P(\epsilon)$ estimate. By definition,
\begin{equation}
    \int_{\rv \in B_{d}(\rv_1)} f(\rv) \, \mathrm{d}\rv \le
    \int_{\rv \in B_{d}(\rv_2)} f(\rv) \, \mathrm{d}\rv.
\end{equation}
Since $\rv_1$ is the MAP estimate,
\begin{equation}
    \int_{\rv \in B_{d}(\rv_2)} f(\rv) \, \mathrm{d}\rv \le
    \int_{\rv \in B_{d}(\rv_2)} f(\rv_1) \, \mathrm{d}\rv.
\end{equation}
Combining the above inequalities together, we conclude
\begin{align}
    (f(\rv_1)-\delta')\int_{\rv \in B_{d}(\rv_1)} \mathrm{d}\rv
     & \le P_1(d)
     \le P_2(d) \le
    f(\rv_1) \int_{\rv \in B_{d}(\rv_2)} \mathrm{d}\rv.
\end{align}
Note that the difference between the upper and lower bounds in the above
inequality is at most $\delta$. Consequently, the difference $\lvert P_1(d)
- P_2(d) \rvert$ is also bounded by $\delta$,
\begin{equation}
    \lvert P_1(d) - P_2(d) \rvert \le \delta \quad \forall \,
    0<d<\epsilon.
\end{equation}
\qed{}

\section{Proof of Theorem~\lowercase{\ref{thm:dom}}}
Recall that the domain of the cost function is given by $\mathcal{Q} \subseteq
\mathbb{R}_{\ge 0}$. For any element $p$ in the range of the cost
function, the inverse image of $p$ is given by
\begin{equation}
    g^{-1}(p) = \left\{ d \in \mathcal{Q} \mid g(d) = p \right\}.
\end{equation}
Let $h(p) = \inf\left( g^{-1}(p) \right)$. From the monotonicity of
$g(d)$, we conclude that $h(p)$ is monotonically increasing in $p$.
Moreover, we have the following relation for any algorithm~$A \in
\mathcal{A}$
\begin{equation}
    P\left( g(D_A) \le p \right) = P\left( D_A \le h(p) \right).
\end{equation}
This allows us to characterize the CDF of $g(D_A)$ as
$F_{A}\left(h\left(g(d_A)\right)\right)$. Moreover, since cost
functions are non-negative, $g(D_A) \ge 0$. Thus the expected value
of $g(D_A)$ for any algorithm $A$ may be expressed as
\begin{equation}\label{eq:expgda}
\E\left[g(D_A)\right] = \int_0^{\sup(\mathcal{Q})} \left(1 -
F_{A}\left(h\left(g(d_A)\right)\right)\right)\, \mathrm{d}d_A.
\end{equation}
Consider any two algorithms $A_1$ and $A_2$ such that $A_1$ stochastically
dominates $A_2$. From~\eqref{eq:sd}, we have
\begin{equation}\label{eq:sd12}
    F_{A_1}\left(h\left(g(d)\right)\right) \ge F_{A_2}\left(h\left(g(d)\right)\right)
    \quad \forall d \in \mathcal{Q}.
\end{equation}
From~\eqref{eq:expgda} and~\eqref{eq:sd12}, we conclude
\begin{equation}
    \E\left[ g(D_{A_1})\right] \le \E\left[g(D_{A_2})\right],
\end{equation}
which proves~\eqref{eq:thmdom}. If $A_1$ strictly dominates $A_2$, then in
addition to~\eqref{eq:sd12}, $A_1$ and $A_2$ satisfies
\begin{equation}\label{eq:ssd12}
    F_{A_1}\left(h\left(g(d)\right)\right) > F_{A_2}\left(h\left(g(d)\right)\right)
    \quad \forall d \in [d_1, d_2].
\end{equation}
From~\eqref{eq:expgda},~\eqref{eq:sd12} and~\eqref{eq:ssd12} we have
\begin{equation}
    \E\left[ g(D_{A_1}) \right] < \E\left[g(D_{A_2})\right].
\end{equation}
\qed{}

\section{Proof of Theorem~\lowercase{\ref{thm:aopt}}}
Say $A \in \mathcal{A}$ is the optimal algorithm for the distance based cost
function $g \in \mathcal{G}$. Further assume that for some algorithm $B \in \mathcal{A}$,
\begin{equation}
    F_A(d) \le F_B(d) \quad \forall d \in \mathcal{Q},
\end{equation}
and that there exists some $d' \in \mathcal{Q}$ such that $F_A(d') < F_B(d')$.
If these conditions implied that $A$ is strictly dominated by $B$,
 then by Theorem~\ref{thm:dom} we have
\begin{equation}\label{eq:BdomA}
    \E\left[ g(D_{A})\right] > \E\left[g(D_{B})\right].
\end{equation}
However, from the optimality of $A$ with respect to $g$,
\begin{equation}\label{eq:AdomB}
    \E\left[ g(D_{A})\right] \le \E\left[g(D_{B})\right],
\end{equation}
which contradicts~\eqref{eq:BdomA}.
To complete the proof of Theorem~\ref{thm:aopt} we need to show the strict
dominance of $B$ over $A$. Since the error CDFs are right continuous, for
any $\epsilon > 0$ there exists a $\delta > 0$ such that
\begin{align}
    F_B(d) - F_B(d') &< \epsilon \quad \forall d \in [d', d'+\delta],
\end{align}
and
\begin{align}
    F_A(d) - F_A(d') &< \epsilon \quad \forall d \in [d', d'+\delta].
\end{align}
Choosing $\epsilon < \frac{F_B(d') - F_A(d')}{2}$ implies
\begin{equation}
    F_B(d) > F_A(d) \quad \forall d \in [d', d'+\delta],
\end{equation}
which proves that $B$ strictly dominates $A$.
\qed{}

\section{Proof of Theorem~\lowercase{\ref{thm:nodom}}}
Since $A_2$ does not stochastically dominate $A_1$, there exists some $d_1 \in \mathcal{Q}$
such that
\begin{equation} \label{eq:f1gf2}
    F_{A_1}\left( d_1 \right) > F_{A_2}\left( d_1 \right).
\end{equation}
Since $A_1$ does not stochastically dominate $A_2$, there exists some $d_2 \in \mathcal{Q}$
such that
\begin{equation} \label{eq:f2gf1}
    F_{A_1}\left( d_2 \right) < F_{A_2}\left( d_2 \right).
\end{equation}
Define the cost functions $g_1$ and $g_{2}$ as
\begin{equation} \label{eq:g1def}
    g_{1}\left(D\right) =
    \begin{cases}
        0 & \text{if } D \leq d_1 \\
        1 & \text{if } D > d_1,
    \end{cases}
\end{equation}
and
\begin{equation} \label{eq:g2def}
    g_{2}\left(D\right) =
    \begin{cases}
        0 & \text{if } D \leq d_2 \\
        1 & \text{if } D > d_2.
    \end{cases}
\end{equation}
Thus,
\begin{align}
    \E\left[ g_{1}(D_{A_1}) \right] &= 1 - F_{A_1} (d_1) \\
    &< 1 - F_{A_2} (d_1)
    = \E\left[ g_{1}(D_{A_2}) \right],
\end{align}
where the inequality follows from~\eqref{eq:f1gf2}.
Similarly,
 \begin{align}
     \E\left[ g_{2}(D_{A_2}) \right] &= 1 - F_{A_2} (d_2) \\
     &< 1 - F_{A_1} (d_2)
     = \E\left[ g_{2}(D_{A_1}) \right],
 \end{align}
where the inequality in the second step follows from~\eqref{eq:f2gf1}.
\qed{}

\section{Proof of Theorem~\lowercase{\ref{thm:area}}}
As $F^*$ dominates every other algorithm, the area under the CDF curve
of any other algorithm is no greater than the area under $F^*$.
Consequently, maximizing $R_A$ is equivalent to maximizing the area under
the CDF curve of the algorithm.  We use this fact to show that the
expected distance error of algorithm $A$, $\E\left[ D_A \right]$
is a linear function of $R_A$.  Thus an algorithm that minimizes $R_A$
minimizes $\E\left[ D_A \right]$ as well and vice versa.  More
formally, for all $A \in \mathcal{A}$
\begin{align}
    R_A &= \int_0^{d^*} \left( F^*(x) - F_A(x) \right) \mathrm{d}x\\
        &= \int_0^{d^*} \left( 1 - F_A(x) \right) \mathrm{d}x +
           \int_0^{d^*} \left( F^*(x)-1 \right) \mathrm{d}x \\
        &= \E\left[ D_A \right] + \alpha,
\end{align}
where the last equality follows from the fact that the random variable
$D_A$ is non-negative and $\alpha$ is the constant given by
$\alpha = \int_0^{d^*} \left( F^*(x)-1 \right) \mathrm{d}x$.
\qed{}

\section{Attainability of the optimal error CDF}\label{apx:attain}
As indicated in Section~\ref{sec:upperboundcomp}, if we have a symmetric
unimodal distribution over a circular area with the mode located at the
center, then the MAP estimate is optimal. On the other hand, from an
algorithmic perspective, MEDE has the nice property, following
Theorem~\ref{thm:aopt}, that if there exists an algorithm that
attains $F^*$, then the MEDE algorithm will attain it as well.
However, it is not clear \emph{a priori} if $F^*$ is attainable. A
sufficient and necessary condition for the attainability $F^*$ may be
obtained in light of the following observations:
\begin{itemize}
    \item By design, the M$P(d)$ estimate matches the performance of
        $F^*$ for the specific distance $d \in \mathcal{Q}$. However,
        there are no guarantees on the performance the estimate at other
        distances.
    \item The M$P(d)$ algorithm need not return a unique location
        estimate. This reflects the fact that there may be multiple
        locations in $\mathcal{S}$ that maximizes the $P(d)$ metric for
        the given posterior distribution.
\end{itemize}
Consequently, consider a modified version of the M$P(d)$ algorithm that
enumerates all possible M$P(d)$ estimates for the given distance $d$. If
$F^*$ is attainable by some algorithm $A$, then the estimate returned by
$A$, say $\hat{\rv}_A$, must exist in the list returned by the M$P(d)$
algorithm. Note that this condition holds for \emph{any} distance $d \in
\mathcal{Q}$ used for the M$P(d)$ algorithm. Consequently, if the
intersection of the estimate list returned by M$P(d)$ for \emph{all} $d
\in \mathcal{Q}$ is non-empty, then we conclude that $F^*$ is attained
by each estimate in the intersection. If the intersection is a null set,
then we conclude that $F^*$ is unattainable, thereby giving us the
necessary and sufficient condition for the existence of $F^*$.

While the above test calls for the repeated execution of the M$P(d)$
algorithm over a potentially large search space, we may exploit the fact
that each execution is independent of each other, thereby gaining
considerable speed improvements by utilizing parallel processing. In
some cases, it may be possible to carry out this test analytically. The
condition enumerated in Section~\ref{sec:upperboundcomp} is a case in
point. The task of identifying the cases when we can obtain the
algorithm that attains $F^*$ is a topic of future research. A slight
generalization of the example given in Section~\ref{sec:upperboundcomp}
leads to the following conjecture.

\begin{con}~\label{rem:bestMLE}
    If the posterior distribution $f_{\Rv}( \rv \vert \Ov = \ov)$ is
    continuous, symmetric and unimodal with the mode located at the
    centroid of a convex space $\mathcal{S}$, then the MAP estimate is
    optimal.
\end{con}

Let $d'$ be the maximal radius of a neighbourhood $N_d(\cv) \subseteq
\mathcal{S}$ centered on the centroid~($\cv$) of our space
$\mathcal{S}$. For all $d \le d'$
\begin{equation}
    \inf_{\rv \in N_d(\cv)} f_{\Rv}( \rv \vert \Ov = \ov) \ge
    \sup_{\rv \in N_d(\rv')\setminus N_d(\cv)} f_{\Rv}( \rv \vert \Ov = \ov),
\end{equation}
for any $\rv' \in \mathcal{S}$. Note that this follows directly from the
assumption that the posterior is unimodal. Consequently, for all $d \le
d'$ the MAP estimate is one of the best performing estimates.
The optimality of MAP at distances beyond $d'$ follows immediately if
\begin{equation}
    \int_{\mathcal{S} \cap (N_d(\cv)\setminus N_d(\rv'))}\mathrm{d}\rv \ge
    \int_{\mathcal{S} \cap (N_d(\rv')\setminus N_d(\cv))}\mathrm{d}\rv,
\end{equation}
for any $\rv' \in \mathcal{S}$. We intuit that the above condition holds
since $\cv$ is the centroid of $\mathcal{S}$, but a proof remains
elusive.

\balance%

\bibliographystyle{IEEEtran}
\bibliography{IEEEabrv,refs}

\end{document}